\documentclass[11pt,a4paper]{article}
\usepackage{jheppub_modified}
\usepackage{amsthm}
\usepackage[T1]{fontenc}
\usepackage[utf8]{inputenc}
\usepackage{mathtools}   
\usepackage{amsfonts}
\usepackage{physics}
\usepackage{mathrsfs}
\usepackage{natbib}
\usepackage{tikz} 
\usepackage{enumitem}
\usepackage{subcaption}
\usepackage{graphicx}

\title{An algebraic study of parametric Stokes phenomena}

\author{In\^{e}s Aniceto$^{1,3,\mathfrak{S}}$ and Samuel Crew$^{2,3,\mathfrak{T}}$}

\affiliation{$^1$School of Mathematical Sciences, University of Southampton, Southampton SO17 1BJ, UK}

\affiliation{$^2$Department of Mathematics, Imperial College London, London SW7 1NE, UK}

\affiliation{$^3$Theoretical Sciences Visiting Program (TSVP), Okinawa Institute of Science and Technology Graduate University, Onna, 904-0495, Japan}

\affiliation{$^{\mathfrak{S}}\mathrm{\textcolor{blue}{I.Aniceto@soton.ac.uk}}$, $^{\mathfrak{T}}\mathrm{\textcolor{blue}{Samuel.c.crew@gmail.com}}$}

\abstract{We investigate geometric aspects of co-equational parametric resurgence, by studying physical problems whose formal asymptotic solutions give rise to Borel transforms lying on an algebraic curve. This perspective allows us to elucidate concepts unique to parametric resurgence such as singularity structures, (virtual) turning points and the higher-order Stokes phenomenon. We construct examples as solutions to Borel plane partial differential equations using an algebraic curve ansatz before turning to the general analytic structure of co-equational resurgence problems, where we provide a systematic description of analytic continuation and Stokes constants through a Borel plane inner-outer matching procedure.}

\theoremstyle{definition}

\newtheorem{example}{Example}[section]
\newtheorem*{remark}{Remark}

\begin{document}
\maketitle

\section{Introduction}

In many areas of theoretical physics and applied mathematics, observables depend on a variety of parameters and understanding their behaviour in different regimes of parameter space is an important problem. Asymptotic analysis \cite{dingle1973asymptotic} with respect to these parameters can play a key role in revealing the underlying analytic structure of such observables. 

Asymptotic expansions of observables often give rise to factorially divergent series, whose sum may be interpreted using Borel resummation whereby the series is realised as the Laplace transform of a locally holomorphic function called the Borel transform. The Borel transform encodes essential information about the original series via its singularity structure and the analytic behaviour of the (resummed) physical observables is intricately linked to the singularities in the Borel plane. In particular, when singularities cross the integration contour in the Laplace transform the Stokes phenomenon is realised and exponentially small corrections---typically invisible in a naïve perturbative expansion---are introduced. Moving in parameter space, these exponentially small correction can grow to dominate the observable's behaviour.

These exponentially suppressed terms form part of a more comprehensive expansion, called a transseries, which incorporates both perturbative and non-perturbative contributions. The study of transseries and the corresponding resurgence structure offers a powerful framework for fully understanding the non-perturbative properties of physical observables. In regimes where the analytical continuation enters regions dominated by these exponentially small terms, their impact becomes critical, and often signals new physical phenomena such as phase transitions.

The theory of Borel resummation and exponential asymptotics has been studied theoretically from a broad range of perspectives, including hyperasymptotics \cite{berry1990hyperasymptotics,berry1991asymptotics,berry1991hyperasymptotics,howls1997hyperasymptotics,olde2005hyperasymptotics, olde2005hyperasymptotics2,boyd1999devil}, \'{E}calle's resurgence \cite{ecalle1977analogue,mitschi2016divergent,sauzin2023resurgence, fauvet2017resurgence} and exact WKB analysis \cite{iwaki2014exact,kawai2005algebraic,honda2015virtual,kidwai2023quantum,nikolaev2024gevrey,nikolaev2023existence,nikolaev2023exact,nemes2025borel,nemes2021borel}. Borel resummation has also proved to be an invaluable tool across a remarkably wide array of physical applications. Examples range from diverse areas of application such as boundary integral problems in water waves and hydrodynamics (\textit{e.g. }\cite{LUSTRI_MCCUE_BINDER_2012,Andersen_Lustri_McCue_Trinh_2024,shelton2024exponential,shelton2024model,Lustri_2022,shelton2023exponential}) to correlation functions in quantum field theory and string theory (\textit{e.g.} \cite{10.21468/SciPostPhys.15.4.179,schiappa2023all,costin2023going,bajnok2023full,fauvet2019holonomy,fujimori2021quantum}). For a more comprehensive review and references on both theory and applications of resurgence we refer the reader to \textit{e.g.} \cite{Aniceto:2018bis} and references therein.

Resurgent analysis is often applied to the asymptotic expansion of large variables of the underlying equation itself, a process often encountered in theoretical physics, and referred to as equational resurgence. In many cases, however, the expansion is performed with respect to an small external parameter (such as $\hbar$ 
in the Schrödinger equation), while the variable of the equation remains finite. This leads to what is known as co-equational resurgence, which is particularly relevant in singularly perturbed problems. The interplay between equational and co-equational resurgence forms the foundation of parametric resurgence \cite{ecalle2002tale,ecalle2022resurgent,ecalletour}. In the co-equational case the singularities in the Borel transform `move' as a function of the equation variables, giving rise to parametric Stokes phenomena, and one uncovers new effects such as the higher-order Stokes phenomena \cite{chapman2007shock,howls2004higher,chapman2005exponential,daalhuis2004higher,howls2012exponentially,nemes2022dingle,berk1982new,howls2024smoothing}.

\paragraph{Background and summary.}

The present work investigates geometric aspects of parametric resurgence from the co-equational perspective. In particular, we will focus on asymptotic problems whose associated Borel transform analytically continues to an algebraic curve. We thus study Borel germs that, on the one hand, satisfy partial differential equations arising from singularly perturbed differential equations, and on the other hand satisfy algebraic polynomial relations. This geometric setting provides a concrete and powerful framework to shed light on parametric Stokes phenomena.  

On the differential equation side, we focus on the case of linear resurgence arising from linear ODEs with an external parameter. We will consider problems of the form:
\begin{equation}
    \label{eq:lin-nth-ODE}
\sum_{i=0}^{d}\,\epsilon^i\,P_i(z,\epsilon)\,\frac{d^i\,y(z)}{dz^i}=\epsilon \,R(z,\epsilon),
\end{equation}
where we assume for simplicity that $P_d(z,\epsilon)=1$ and the other $P_i(z,\epsilon),\,R(z,\epsilon)$, $i=0,\cdots,d-1$ are meromorphic (in fact polynomial) in both $z$ and $\epsilon$. $R(z)$ is a meromorphic in-homogeneous term. We study formal parametric transseries solutions to these equations of the form:
\begin{eqnarray}
\label{eq:transs-sol-nth}
    y(z,\epsilon) & = & \sum_{i=0}^{d} \sigma_i(\epsilon)\,\mathrm{e}^{-\chi_i(z)/\epsilon}\, y^{(i)}(z,\epsilon)\\
    & = & \sigma_0(\epsilon)\,\mathrm{e}^{-\chi_0(z)/\epsilon}\,(\epsilon\,y^{(0)}_0(z) + \epsilon^2 y^{(0)}_1(z) + ...) + \sigma_1(\epsilon) \,\mathrm{e}^{-\chi_1(z)/\epsilon}(y_0^{(1)}(z) + \ldots) + \ldots\nonumber
\end{eqnarray}
where all the terms are meromorphic functions in $z$ and $\chi_0(z) = 0$. The transseries sectors $y^{(i)}(z,\epsilon)$ should be seen as formal expansions in small $\epsilon$ with $z-$dependent coefficients. These expansions will generally be Gevrey-1 asymptotic in $\epsilon$ (zero radius of convergence in $\epsilon$ for fixed $z$ stemming from the factorial growth of the coefficients), and this transseries will have parametric  resurgent properties. The coefficients $\sigma_i(\epsilon)$ are fixed by some initial/boundary conditions, with $\sigma_0(\epsilon)$ being fixed by the particular solution of the ODE. Note that the series $y^{(0)}$ starts at order $\epsilon$ for convenience. If this were not the case in particular examples, this could be absorbed in the definitions of $\sigma_0(\epsilon)$. We have chosen this particular form of the ODEs for simplicity, but one could also consider more general linear ODEs, with a more general form of transseries.

The usual theory of Borel analysis and resummation associates a parametric holomorphic function $\varphi_B(w,z)$ to such transseries, with isolated singularities corresponding to exponential weights $\chi_i(z)$ in the transseries, also known as \textit{singulants}. The $y^{(0)}(z,\epsilon)$ sector of the transseries is completely fixed by the inhomogeneous term of the differential equation and we may take the Borel transform to obtain the parametric germ
\begin{equation}
\label{eq:borel-germ-at-zero}
\varphi (w,z)\equiv\,\sum_{n=0}^{\infty}\varphi^{(0)}_n(z)w^n=\,\sum_{n=0}^{\infty}\,\frac{y^{(0)}_n(z)}{\Gamma(n+1)}\,w^{n}\,.
\end{equation}
The re-summed result for the formal series is then given by the Laplace transform in the complex plane
\begin{equation}
    \mathcal{S}y^{(0)}(z,\epsilon)=\int_{0}^{\infty}\mathrm{d}w\,\mathrm{e}^{-w/\epsilon}\,\varphi(w,z).
\end{equation}
Generally we may consider an arbitrary direction $\theta$, but since we consider $\epsilon > 0$ throughout we may without loss of generality take $\theta =0$.

As we vary $z \in \mathbb{C}_z$, singularities in the analytic continuation of $\varphi(w,z)$ associated to transseries components $w=\chi(z)$ may cross the integration contour. This happens along real dimension one loci $l \subset \mathbb{C}_z$ known as Stokes lines. The summations on either side of such a line will differ by some exponentially small contribution. This is the well-known Stokes phenomenon. In the transseries language, this corresponds to the parameters $\sigma_i(\epsilon)$ jumping by a known (or calculable) value, the Stokes constants --- described by the so-called Stokes automorphism that acts on the transseries.

Interestingly, when directly seeking a formal parametric transseries ansatz solution to \eqref{eq:lin-nth-ODE} we find that while the $y^{(0)}(z,\epsilon)$ sector is completely fixed by the inhomogeneous term, the higher components $y^{(i)}_k(z)$ solve a sequence of linear differential equations in $z$. The initial data for these ODEs, say the formal series $y^{(i)}(z_{\star},\epsilon)$ at some point $z_\star$, are undetermined since they may be equivalently absorbed into the $\epsilon$-dependent constants $\sigma_i(\epsilon)$. The exponential factors $\chi_i(z)$ also solve an algebraic-differential equation with similarly undetermined constants, namely the zeroes $\chi(z)=0$ of the singulants. One of the goals of this work is to understand these ambiguities and seemingly undetermined constants. In section \ref{sec:Inner-outer} we outline an analytic continuation method of the Borel germ across its various singularities that allows us to unambiguously fix these constants and consequently determine the initial data for the ODEs satisfied by the transseries sectors. The technique relies on the coalescence of Borel singularities (coalescence is an important phenomena in varying asymptotic context see \textit{e.g.} \cite{berry1993unfolding,berry1994overlapping}) and Darboux's theorem, and is equivalent to inner-outer matched asymptotics in the physical plane. We show that the Borel approach provides a fully systematic way to obtain an \textit{unambiguous} transseries solution in co-equational resurgence, while also allowing us to understand the $\epsilon$-dependence of the Stokes constants.

In parallel, we begin this work in section \ref{sec:algcurves} with an abstract discussion of simple examples of parametric algebraic curves $\Sigma_z$. That is loci in $\mathbb{C}^2$ cut out by a single $z$-dependent polynomial
\begin{equation}
    \Sigma_z \subset \mathbb{C}_w \times \mathbb{C}_{\varphi}\, : \, P_z(w,\varphi) = 0\,.
\end{equation}
Such algebraic curves have a finite number of branch points with $z$-dependent locations $\{\chi(z)\}$. Given a complex parameter $\epsilon$ and a choice of Hankel contours around the singularities we discuss in section \ref{sec:stokes} how to associate a natural transseries structure, with Stokes automorphisms, to $\Sigma_z$. The goal of this work is to elucidate aspects of co-equational resurgence by studying examples where these abstract structures coincide with those originating from formal transseries solutions to singularly perturbed ODEs. To this end, we re-write the physical problem \eqref{eq:lin-nth-ODE} as a PDE $\mathscr{P}_B$ for the Borel germ and describe a method that allows us to construct example pairs of Borel PDEs and algebraic curves.

Parametric resurgent problems have interesting additional features known as virtual turning points \cite{honda2015virtual} and the higher-order Stokes phenomenon. The higher-order Stokes phenomenon occurs when the analytic continuation has a branched structure causing the intersection conditions of singularities and contours in the Borel plane to be different from the na\"{i}ve expectation from the dominance conditions of the transseries solution. Namely, na\"{i}ve Stokes lines can be illusory. In section \ref{sec:stokes} we associate an additional transseries structure from the large $n$ asymptotics of (singular) parametric germ coefficients $\varphi_n(z)$ associated to $\Sigma_z$. We find a conformal relationship between the two types of transseries -- the physical, small $\epsilon$ expansion \eqref{eq:transs-sol-nth} and the large $n$ coefficient expansion -- and show that the higher-order Stokes phenomena in the $\epsilon$ transseries has its origin in the ordinary Stokes phenomena in the $1/n$ transseries. We elucidate the co-equational resurgent structure in terms of the composition of Stokes automorphism corresponding to both the $\epsilon$ and the $1/n$ transseries along oriented paths in $\mathbb{C}_z$. 

The advantage of the algebraic approach summarised above is that it allows us to construct a powerful collection of examples to study every aspect of parametric Stokes phenomena, including the higher-order Stokes phenomena, explicitly and clearly understand their geometric origin.

\paragraph{Outline.}
This paper is organised as follows. We begin in section \ref{sec:algcurves} with a discussion of parametric algebraic curves and the singularity structures that will be relevant for co-equational resurgence. We discuss how to construct physical problems whose formal solutions give rise to Borel germs that lie on such algebraic curves. In section \ref{sec:stokes}, we turn to the Stokes phenomena and discuss how to associate transseries structures to algebraic curves. In this section we also elucidate the geometric origin of the higher-order Stokes phenomena and the relationship between small $\epsilon$ transseries and large $n$ transseries of singular germ coefficients. Finally, in section \ref{sec:Inner-outer} we discuss how to recover these structures from a formal parametric transseries ansatz to a physical singularly perturbed ODE through a germ inner-outer matching procedure.

\section{Algebraic curves and the Borel transform}\label{sec:algcurves}

Singularly perturbed differential equations often admit formal solutions given by parametric perturbative series. Considering the Borel transform of such perturbative series we are led to consider a parametric germ on $\mathbb{C}_w$ with local holomorphic dependence on a parameter $z \in \mathbb{C}_z$. We are interested in cases where, in addition, the analytic continuation of this germ lies on an algebraic curve. This leads to a geometric perspective which sheds new light on familiar concepts in singular perturbation theory, such as virtual turning points. In this section we discuss the general singularity structures involved and introduce a method to generate examples.

\subsection{Algebraic curves}
\begin{figure}
    \centering
    \includegraphics[scale=0.23]{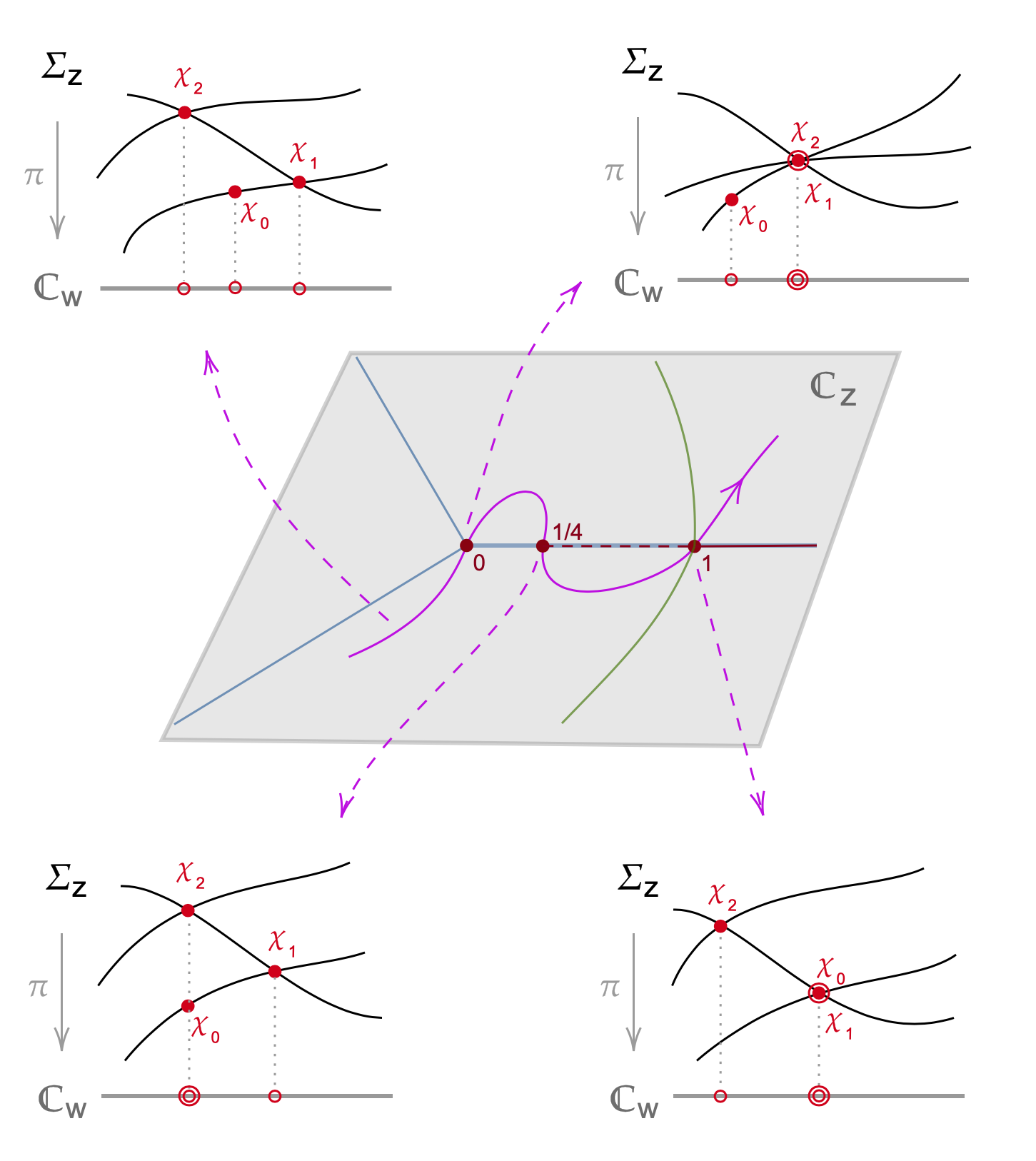}
    \caption{A visualisation of $\Sigma_z$ with parametric dependence on $z$. In the top-left panel we show the sheets at a generic $z$. Bottom-left are the sheets at $z=1$ where we have a $(0,1)$ turning point. Bottom-right are the sheets at $z=0$ where we have a $(1,2)$ turning point. Top-right is the configuration at the virtual turning point where $\pi \chi_0(z) = \pi \chi_2(z)$ but $ \chi_0(z) \neq \chi_2(z)$.}
    \label{fig:germexample}
\end{figure}

We consider Gevrey-1 formal series $\mathbb{C}[[\epsilon]]_1(z)$ whose Borel transforms lead to parametric germs in $\mathbb{C}\{w\}(z)$. We denote the (parametric) analytic continuation of the germ by $\varphi(w;z)$ and we are concerned with examples where the locus
\begin{equation}
    \Sigma_z = \{(w,\varphi(w;z))\, :\, w\in \mathbb{C}_w\} \subset \mathbb{C}_w \times \mathbb{C}_{\varphi}\,,
\end{equation}
is an algebraic curve defined by the zero set of a $z$-dependent polynomial $P_z(w,\varphi) = 0$. We also further restrict to examples where $\Sigma_z$ is single-valued in $z \in \mathbb{C}_z$ throughout. In particular, the coefficients of $P_z(w,\varphi)$ are meromorphic functions of $z \in \mathbb{C}_z$.  The projection $\pi: \Sigma_z \to \mathbb{C}_w$ may be branched and, additionally, $\varphi(w;z)$, considered as a function\footnote{That is $\varphi(w;z) := \varphi(\pi^{-1}(w),z)$ for a particular choice of pre-image.} on $\mathbb{C}_w$, may have poles. Since the polynomial $P_z(w,\varphi)$ depends parametrically on $z$, the location of branch points and poles varies with $z$. We provide an illustrative sketch of this setup in figure \ref{fig:germexample}.

\begin{remark}
We will later see that the choice of an initial parametric germ/Gevery-1 formal series distinguishes a pre-image $\mathbf{0}=\pi^{-1}(0) \in \Sigma_z$. We will see momentarily that for curves arising as perturbative solutions to inhomogeneous linear problems, this point will be regular whereas for solutions to homogeneous problems it may be itself a branch point.
\end{remark}

\begin{example}\label{eg:runningexample}
Throughout the work, we will consider the following algebraic curve as a running example
\begin{equation}
    \Sigma_z = \{ (w,\varphi(w,z)) \, : \, P_z(w,\varphi) = 0 \} \subset \mathbb{C}_w \times \mathbb{C}_{\varphi}\,,
\end{equation}
with 
\begin{equation}
    \label{eq:curve-main-example}
    P_z(w;\varphi) = 4(w-\chi_1(z))(w-\chi_2(z))\,\varphi^2 - 4(w-\chi_1(z))\,\varphi + 1,
\end{equation}
and
\begin{equation}\label{eq:ex1-branch-point-pole}
    \chi_1(z) = z^2 / 2\quad;\quad \chi_2(z) = z-1/2\,.
\end{equation}
The Riemann surface is sketched in figure \ref{fig:Hankeleg}. As we progress we will add more detail to this example: In the present subsection, we will understand various singularities associated to $\Sigma_z$ and local germ expansions. In the following subsection \ref{subsec:curveansatz} we will discuss how this curve arises as the Borel solution to a singularly perturbed differential equation. In section \ref{sec:stokes} we will understand the Stokes phenomena associated to the geometry of $\Sigma_z$ before finally in section \ref{sec:Inner-outer} introducing a complex analytic inner-outer matching procedure to reconstruct the algebraic curve from transseries asymptotics.
\end{example}

\subsection{Singularity structures}
We now consider certain singularities associated to the parametric curve $\Sigma_z$. We will later interpret these singularities in the context of matched asymptotics, but for now they are solely geometric properties of the curve.

\paragraph{Borel singularities $\Gamma_w(z)$.}
We first define a subset $\Gamma_w(z)$ of $\Sigma_z$ consisting of the branch points of $(\Sigma_z,\pi)$ and poles of $\varphi_B(\pi^{-1}(w),z)$. The set $\Gamma_w(z)$ is thus a set of pairs of points $(\chi(z),\varphi_B(\chi(z);z))$. We will also often consider the projection of $\Gamma_w(z)$ to $\mathbb{C}_w$, denoted $\hat{\Gamma}_w(z) := \pi \circ \Gamma_w(z)$. In a slight abuse of notation (particularly when discussing formal transseries), 
we will often write elements of this projection as $\chi(z)$, even though $\chi(z)$ is more properly a point of $\Sigma_z$. The elements of the set $\hat{\Gamma}_w(z)$ will later be associated with the exponential weights of transseries components \textit{i.e.} $\chi(z)$ are singulants. 

The singularity set $\hat{\Gamma}_w(z)$ may be obtained by studying the discriminant of $P_{z}(w,\varphi)$:
\begin{equation}
    \Delta_{P_z}(w) = \sum_{i,j}(\varphi_i(w;z)-\varphi_j(w;z))^2,
\end{equation}
where $\varphi_i$ are roots of $P_z(\varphi_i,w)=0$. When $\Delta_P = 0$, the projection $\pi: \Sigma_z \to \mathbb{C}_w$ has a ramification/branch point.

\paragraph{Physical singularities $\Gamma_z^{(\chi)}$.}
We shall now turn to a second type of singularity. Local germ expansions of algebraic curves have branch point singularities with  rational order $\alpha$. To each branch point or pole $\chi(z)$ in $\Gamma_w(z)$ we may then consider a singular parametric germ expansion:
\begin{equation} \label{eq:Borel-germ-at-chi-i}
    \varphi_B(w,z) \equiv \Phi^{(\chi)}(w,z) =  
    \left(w-\chi(z)\right)^{-\alpha_{\chi}}\sum_{k\ge0}\,\Phi_{k}^{(\chi)}(z)\,(w-\chi(z))^{k}+\mathrm{regular}\,.
\end{equation}
We then define a second singularity set, denoted $\Gamma^{\chi}_z$, as a subset\footnote{Recall that we assume $\Sigma_z$ is single-valued in $\mathbb{C}_z$.} of $\mathbb{C}_z$. $\Gamma_z^{\chi}$ is the set of singularities of the coefficients $\{\Phi_n^{\chi}(z)\}$ of this expansion. In our examples the coefficients $\{\Phi_n^{\chi}(z)\}$ will be meromorphic and every $\Gamma_z^{\chi}$ will be a finite set. 

\begin{example}\label{eg:runninggerms}
We now discuss the singularity sets $\Gamma_w(z)\subset \Sigma_z$ and $\Gamma^{\chi}_z \subset \mathbb{C}_z$ associated to our running example \ref{eq:curve-main-example}. Firstly we compute the discriminant of this quadratic curve
\begin{equation}
    \Delta_P = 16(\chi_1(z)-\chi_2(z))(w-\chi_1(z)),
\end{equation}
and we thus find a single\footnote{Assuming we are away from the degenerate point $z=1$. The interpretation of this special point will be discussed later in section \ref{sec:stokes}.} branch point at $w=\chi_1(z)$. We may solve $P_{z}(\varphi,w)=0$ in closed form to find 
\begin{equation}\label{eq:ex1-explicit-2sheeted-sol}
    \varphi=\frac{1}{2(w-\chi_2)}\left(1 \pm \sqrt{1-\frac{w-\chi_2}{w-\chi_1}}\right)\, ,
\end{equation}
where we note that on one sheet of $\pi: \Sigma_z \to \mathbb{C}_w$ we have in addition a simple pole at $w=\chi_2(z)$. We will later consider $\Sigma_z$ as the Borel transform of the asymptotic solution to an \textit{inhomogeneous} linear differential equation, in which case the origin $w=0$ is further distinguished as a relevant (albeit non-singular) Borel germ. We may thus write the singularity set for this example as
\begin{equation}
    \hat{\Gamma}_w(z) = \{ \chi_0(z)=0, \,\, \chi_1(z) = z^2/2,\,\, \chi_2(z) = z-1/2\}\,.
\end{equation}
The lift $\Gamma_w(z)$ to $\Sigma_z$ is clear from the exact solution \eqref{eq:ex1-explicit-2sheeted-sol}.

We can at this point determine the germ expansions $\Phi^{(\chi)}$ of the form \eqref{eq:Borel-germ-at-chi-i} at the points of $\Gamma_w(z)$. To obtain these expansions from the algebraic curve we substitute an ansatz of the form \eqref{eq:Borel-germ-at-chi-i} into the algebraic equation or, in our simple example, invert the defining equation \eqref{eq:curve-main-example} and determine the expansion coefficients.

First, we consider the (non-singular, $\alpha_0 = 0$) germ expansion near $w=\chi_0(z)=0$. At $w=0$ the curve takes the form
\begin{equation}
    P_{z}(w=0,\varphi) = 4\chi_1(z)\chi_2(z) \varphi^2 + 4\chi_1(z) \varphi + 1\,,
\end{equation}
we see thus have two possible germ expansions at $w=0$ with first terms given by
\begin{equation}
    \Phi_0^{(0)} = \frac{1}{z(1-2z)},\,\, \text{or}\quad  \Phi_0^{(0)} = - \frac{1}{z}.
\end{equation}
Both are valid choices of distinguished origin but in later sections we study the inhomogeneous equation with a choice of inhomogeneous term that selects the second possibility and thus proceed with the choice of $\Phi_0^{(0)}=-1/z$. It is straightforward to obtain the remaining coefficients $\Phi^{(0)}_k(z)$ of the germ \eqref{eq:Borel-germ-at-chi-i} at the origin $\Phi^{(0)}$ in closed form as
\begin{equation}\label{eq:coeffs-Phi0-curve}
    \Phi_k^{(0)}=-\frac{2^k}{\sqrt{\pi}\,z}\left(\sum_{n=\lfloor \frac{k+1}{2}\rfloor}^k \frac{\Gamma(n+1)\,\Gamma\left(n+1/2\right)}{\Gamma(2n+1-k)\Gamma(k+1)}\,\frac{1}{z^{2n}}+\sum_{n=\lfloor \frac{k+2}{2}\rfloor}^{k} \frac{\Gamma(n)\,\Gamma\left(n+1/2\right)}{\Gamma(2n-k)\Gamma(k+1)}\,\frac{1}{z^{2n-1}}\right).
\end{equation}

Turning to the coefficients of the singular germ at $\chi_1(z)=z^2/2$, close to the branch point $w=\chi_1$ we have
\begin{equation}
    \Phi^{(1)}(w,z)\sim \frac{1}{\sqrt{w-\chi_1}}\frac{\sqrt{\chi_2-\chi_1}}{2(w-\chi_2)}+\mathrm{reg.}\, ,
    \label{eq:curve-Phi1-germ}
\end{equation}
where we should view the above expression as an expansion like \eqref{eq:Borel-germ-at-chi-i}, in half-integer powers of $(w-\chi_1)$ with $\alpha=1/2$, and the choice of the branch of the square root should be taken to match the analytic continuation of the choice at $w=0$. (Note that this analytic continuation will depend on the complex value of $z$.)  The coefficients of this germ expansion (positive branch) will be given by
\begin{equation}\label{eq:coeffs-Phi1-curve}
    \Phi_k^{(1)}(z)=\frac{(-2)^{k-1/2}}{(1-z)^{2k+1}}\,.
\end{equation}
Finally, at $w=\chi_2$ singular germ expansion can be determined from \eqref{eq:ex1-explicit-2sheeted-sol} and is a simple pole, whose residue will either be one or zero (\textit{i.e.} no pole) depending on the sheet. We will define the germ at this point as 
\begin{equation}
    \Phi^{(2)}(w,z)\sim \frac{1}{w-\chi_2}+\textrm{reg.}
\end{equation}
\end{example}

\paragraph{Turning points.}
A $(i,j)$ \textit{turning point} $z_{ij} \in \mathbb{C}_z$ is where two elements $\chi_i$ and $\chi_j$ in $\Gamma_w(z)$ coincide $\chi_i(z_{ij}) = \chi_j(z_{ij})$ \textit{as points on $\Sigma_z$}. On the other hand, a \textit{virtual turning point} is a point $z_{ij}$ that is not a turning point but where the projection under $\pi$ of two points in $\hat{\Gamma}_w(z)$ coincide, that is $\pi(\chi_i(z_{ij}))=\pi(\chi_j(z_{ij}))$. In contrast with virtual turning points, turning points are associated to the singularity sets $\Gamma^{\chi_i}_z$ since when coefficients of a germ associated to $\chi_i$ diverge, then this is because another singularity $\chi_j$ is coalescing to that point on $\Sigma_z$ and the radius of convergence is shrinking to zero.

\paragraph{Singularity graph $\Gamma(\Sigma_z)$.}
We conclude by noting that the singularity structure of $\Gamma_w(z)$ and $\Gamma^{\chi}_z$ may be encoded by a graph $\Gamma(\Sigma_z)$, an example of which is given in figure \ref{fig:graphexample}. We first draw a node for each element  $\chi \in \Gamma_w(z)$. We then draw a `child' node for each element of the associated $\Gamma^{\chi}_z$. The graph is connected by oriented arrows $i \leftarrow j$ each with a tail at a main node $\chi_j$ and a head at a child node $z_{ij} \in \Gamma^{(i)}_z$ which indicates that $\chi_j$ is coalescing to $\chi_i$ when the parameter $z = z_{ij}$.

\begin{remark}
Note that in examples where $\Sigma_z$ is pole free then the graph $\Gamma$ is symmetric. That is every arrow $i \leftarrow j$ has a partner $i \rightarrow j$. Since in this work we are interested in examples where $\Sigma_z$ may have poles, and with a non-singular distinguished origin, we must also consider oriented graphs.
\end{remark}

\begin{example}
We can now complete our discussion of the singularity structure in our running example. We have already determined that 
\begin{equation}
    \hat{\Gamma}_w(z) = \{ \chi_0(z)=0, \,\,\chi_1(z)=z^2/2, \,\,\chi_2(z)=z-1/2\},
\end{equation}
Considering the germ expansion from section \ref{eg:runninggerms} about each of these points, we can see that the $\mathbb{C}_z$ singularity sets are given by
\begin{equation}\label{eq:egturningpoints}
    \Gamma_z^{(0)} = \{ 0\}, \quad \Gamma_z^{(1)} = \{ 1\}, \quad \Gamma_z^{(2)} = \emptyset.
\end{equation}
We further see a $(0,1)$ turning point at $z=0$, a $(1,2)$ turning point at $z=1$. And $z=1/2$ is in fact a $(0,2)$ \textit{virtual} turning point since at this point clearly $\pi \chi_1(z) = \pi \chi_2(z)$ however we have that $\chi_2(z)$ and $\chi_0(z)$ are on different sheets of $\pi: \Sigma_z \to \mathbb{C}_w$. This data is encoded in the associated singularity graph illustrated on the right hand side of figure \ref{fig:graphexample}. We will revisit the turning point structure again later, summarised in the Stokes graph figure \ref{fig:stokesgrapheg}.
\end{example}

\subsection{Borel PDE and curve ansatz}\label{subsec:curveansatz}
We have so far dealt with the geometry of curves $\Sigma_z$ that arise as the analytic continuation of a parametric holomorphic germ. In this subsection we are interested more specifically in the case that $\varphi_B$ is a germ that arises as the Borel transform of a parametric asymptotic series generated by a formal solution to a singularly perturbed differential equation. Let us thus consider a physical singularly perturbed differential operator of the form 
\begin{equation}
    \mathscr{P} = \epsilon^d \frac{d^d}{dz^d} + P_{d-1}(z) \epsilon^{d-1} \frac{d^{d-1}}{dz^{d-1}} + \ldots + P_{0}(z), 
\end{equation}
and an associated inhomogeneous problem $\mathscr{P}y(z,\epsilon) = \epsilon R(z)$. For simplicity We consider meromorphic $\{P_k(z)\}$. Under suitable conditions \cite{nikolaev2023existence,nemes2025borel}, such equations admit Borel summable Gevery-1 formal asymptotic solutions of the form
\begin{equation}
    y(z,\epsilon) = \epsilon\, y_0(z) + \epsilon^2 \,y_1(z) + \epsilon^3 \,y_2(z) + \ldots
\end{equation}
and we write the corresponding Borel germ \eqref{eq:borel-germ-at-zero} as
\begin{equation}
    \varphi_B(w,z) = \sum_{n=0}^{\infty}\frac{y_n(z)}{n!}w^n.
\end{equation}
In the Borel plane, the corresponding differential problem for the germ $\varphi_B(w,z)$ is given in terms of the linear \textit{partial} differential operator
\begin{equation}
\label{eq:Borel-PDE-gen}
    \mathscr{P}_B = \partial_z^d + P_{d-1}(z) \partial_z^{d-1} \partial_w + \ldots + P_{0}(z) \partial_w^{d} \, ,
\end{equation}
and one should solve $\mathscr{P}_B \varphi_B(w,z) = 0$ with the initial data $\varphi_B(w=0,z) = R(z)$. We are thus interested in parametric holomorphic functions $\varphi_B(w,z)$ that solve such partial differential equations and simultaneously whose analytic continuation (which, in a minor abuse of notation, we again denote as $\varphi_B$) lies on an algebraic curve. In other words, we want to study the simultaneous problems:
\begin{equation}\label{eq:curveandPDE}
    \mathscr{P}_B \varphi_B = 0 ,\,\quad \text{and} \quad P_z(w,\varphi_B(w,z)) = 0 \,,
\end{equation}
with the additional constraint for inhomogeneous problems that  $\varphi_B(w=0,z) = R(z)$ so that $P_z(w=0,R(z)) = 0$. In this subsection we discuss the question of whether such solutions exist and provide a method that allows one to construct pairs of example curves and PDEs satisfying the conditions \eqref{eq:curveandPDE}.

\paragraph{Ansatz.}
The idea is simple, we consider an ansatz for a curve and an ansatz for a PDE problem and solve the resulting consistency equations. Let us begin with an ansatz in the form of a degree $D$ algebraic curve:
\begin{equation}\label{eq:algebraicansatz}
    P_z(w,\varphi) = \sum_{k=0}^D a_k(w,z) \varphi^k
\end{equation}
with $a_k(w,z)$ degree $d_k$ polynomials in $w$ with unspecified functions of $z$ as coefficients $a_k^{(l)}(z)$. Namely,
\begin{equation}
    a_k(w,z) := \sum_{l=0}^{d_k} a_k^{(l)}(z)w^l \,.
\end{equation}
We note that if we substitute a holomorphic germ for $\varphi(w,z)$ about the origin
\begin{equation}
    \label{eq:germ-ansatz-curve}
    \varphi(w,z) = \sum_{n=0}^{\infty}\varphi_n(z)w^n \,,
\end{equation}
then setting $P=0$ in the algebraic equation \eqref{eq:algebraicansatz} is equivalent to specifying an \textit{algebraic nonlinear recursion relation} for the coefficients $\varphi_n(z)$. For example, at first order in $w$ we find the relation
\begin{equation}\label{eq:curverecursion}
    \sum_{k=0}^D a_k^{(0)}(z) \varphi_0(z)^k = 0 \,.
\end{equation}
One may continue to higher orders, obtaining increasingly more complex algebraic relations for the $\varphi_n(z)$. On the other hand, the Borel PDE \eqref{eq:Borel-PDE-gen} is also amenable to a germ ansatz \eqref{eq:germ-ansatz-curve},
and in that case we obtain a \textit{differential linear recurrence relation}, and thereby explicit formulae for the $\varphi_n(z)$ in terms of derivatives of the coefficients $P_i(z)$ and $R(z)$. For example, consider a second order linear ODE:
\begin{equation}
\label{eq:lin-2nd-ODE}
    \epsilon^2 y''(z) + \epsilon P_1(z) y'(z) + P_0(z) y(z) = \epsilon R(z) \,,
\end{equation}
where $P_i(z)$ and $R(z)$ are meromorphic (in fact polynomial). The corresponding Borel PDE takes the form:
\begin{equation}
    \label{eq:Borel-2nd-PDE}
    \partial_z^2\varphi(w,z) + P_1(z)\, \partial_z \partial_w \varphi(w,z) + P_0(z)\,\partial_w^2 \varphi(w,z)=0.
\end{equation}
The germ $\varphi(w,z)$ at $w=0$ is related to the perturbative solution of the differential equation $y^{(0)}(z,\epsilon)$ by the usual relation \eqref{eq:borel-germ-at-zero}. The coefficients $\varphi_{k}^{(0)}$ obey the recursion ODEs:
\begin{equation}
\varphi_{k}^{(0)}(z)=-\frac{1}{k(k-1)}\frac{1}{P_0(z)}\left(\varphi_{k-2}^{(0)''}(z)+(k-1)P_1(z)\,\varphi_{k-1}^{(0)'}(z)\right)\,,
\label{eq:PDErecursion}
\end{equation}
where $\varphi_{0}^{(0)}$ and $\varphi_{1}^{(0)}$ are initial conditions which are not determined by the Borel PDE, but that can be determined from
physical initial data coming from directly from
the inhomogeneous part of the physical ODE
\begin{equation}
\label{eq:leading-coefficients}
\varphi_{0}^{(0)}=\frac{R(z)}{P_0(z)}\,;\;\varphi_{1}^{(0)}=-\frac{P_1(z)}{P_0(z)}\partial_{z}\varphi_{0}^{(0)}.
\end{equation}
We then substitute \eqref{eq:PDErecursion} into \eqref{eq:curverecursion} and consistency of these two recursion relation yields a  set of equations for the functions $a^{(k)}_n(z)$. To construct examples, we search over increasing degrees $D$ and $d_1, d_2, \ldots, d_k$ until a solution is found.

\begin{example}
Consider the case $P_1(z) = 1+z$, $P_0(z) = z$ and $R(z) = -1$ so that the Borel PDE reads
\begin{equation}\label{eq:running-example-PDE}
    \mathscr{P}_B = \partial_z^2 + (1+z)\partial_{z}\partial_w + z \partial^2_w\,.
\end{equation}
We search up to degree $D=2$ and with degrees $d_1 = 2$, $d_2 = 2$ and $d_3 = 2$ we recover the algebraic curve solution of our running example \eqref{eq:curve-main-example}:
\begin{equation}
\begin{split}
    P_z(w,\varphi) &= \left(4 w^2-2 w z^2-4 w z+2 w+2 z^3-z^2\right)\varphi^2 +\left(2 z^2-4 w\right)\varphi +1\,,\\
    &= 4(w-\chi_1(z))(w-\chi_2(z))\varphi^2 - 4(w-\chi_1(z))\varphi +1 .
\end{split}
\end{equation}
We have thus now realised our running example both as as a solution to a Borel PDE / perturbative asymptotics problem and solving an algebraic curve. The corresponding physical inhomogeneous ODE is:
\begin{equation}\label{eq:running-example-ODE}
    \epsilon^2 y''(z) + \epsilon (1+z) y'(z) + z y(z) = - \epsilon \,.
\end{equation}
The geometry of the curve $\Sigma_z$ governs the asymptotics. This equation was also recently studied in \cite{shelton2023exponential} from the applied exponential asymptotics perspective. 
\end{example}

\begin{figure}
    \centering
    \includegraphics[scale=0.2]{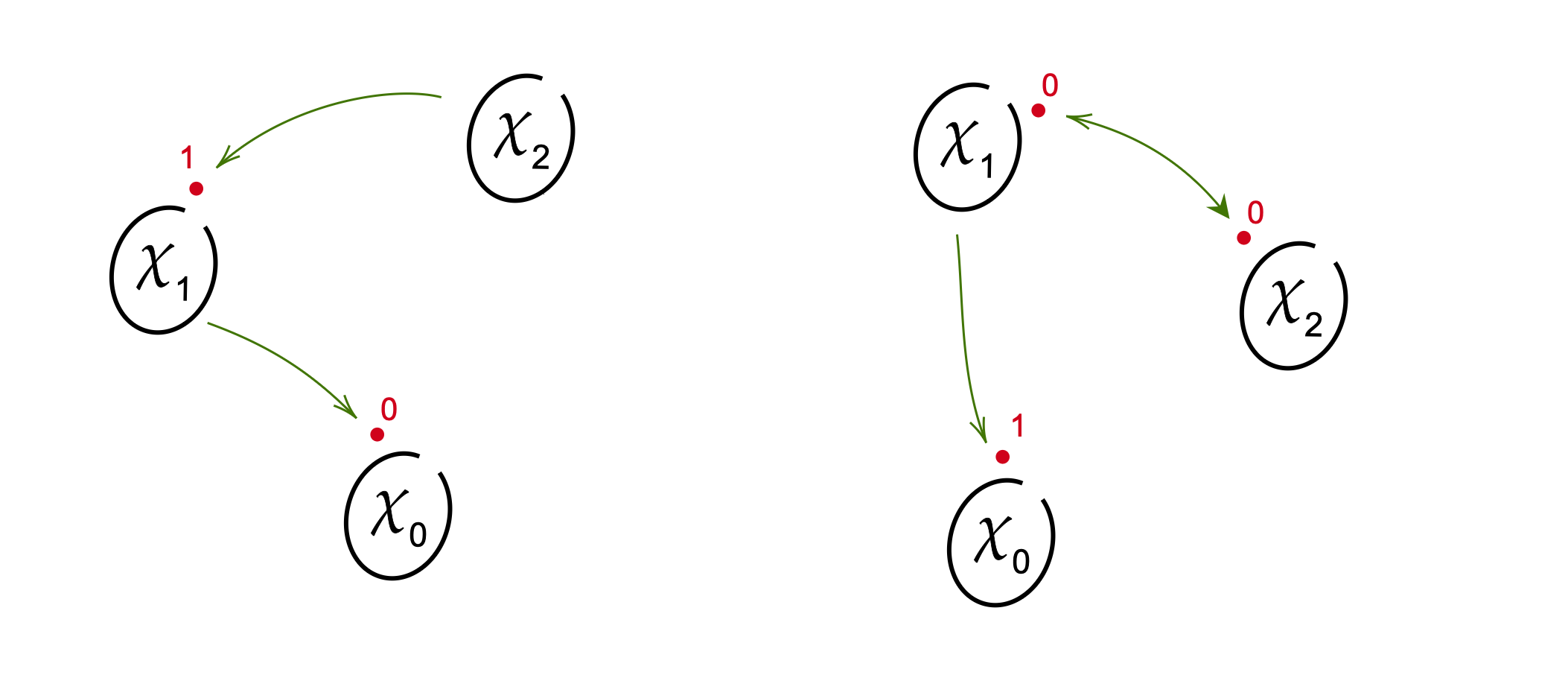}
    \caption{Singularity graphs. The graph on the left is associated to our quadratic running example \eqref{eq:curve-main-example}, while the graph of the right is associated to the second example \eqref{eq:eg2curve}.}
    \label{fig:graphexample}
\end{figure}

\begin{example}\label{eg:second}
Consider the equation
\begin{equation}\label{eq:eg2phys}
    \epsilon^2 y''(z) + 2\epsilon y'(z) + (1-z) y(z) = \epsilon\,,
\end{equation}
Following the algorithm above, we find now an associated cubic curve in this case:
\begin{equation}\label{eq:eg2curve}
\begin{split}
    &\Sigma_z = \{(w,\varphi) \, :\, P_{z}(w,\varphi) = 0\}, \\
    &P_z(w,\varphi)  = -((-1+6z - 9z^2 + 4z^3)+ 6(-1+3z)w - 9w^2 )\varphi^3 + (3z)\varphi -1\,.
\end{split}    
\end{equation}
We now illustrate the ideas of the present section for this example. We first study the singularity sets. The discriminant of $P_z$ may be written as
\begin{equation}
\begin{split}
    \Delta_{P_z}(w) &= -108 z^3 - 648 w z^3 - 972 w^2 z^3 + 648 z^4 + 1944 w z^4 - 972 z^5 + 
 432 z^6 \\
 &= -972 z^3\left(w-\chi_1(z)\right)\left(w-\chi_2(z)\right),
\end{split} 
\end{equation}
where 
\begin{equation}
    \chi_1(z) = \frac{1}{3}(-1+3z - 2 z^{3/2}),\quad \chi_2(z) = \frac{1}{3}(-1+3z + 2 z^{3/2}).
\end{equation}
Thus setting $\Delta_P=0$ we find branch point singularities at $\chi_1(z)$ and $\chi_2(z)$. Again, since we consider here $\Sigma_z$ as a solution to an inhomogeneous problem, the origin $\chi_0(z) = 0$ is also distinguished. We see that
\begin{equation}
    P_{z}(w=0,\varphi) = -1 + 3 z \varphi -  (-1 + 6 z - 9 z^2 + 4 z^3)\varphi^3 ,
\end{equation}
And choose the origin corresponding to the branch
\begin{equation}
    \varphi^{(0)}_0 = \frac{1}{1-z},
\end{equation}
in order to match the perturbative solution to the physical problem \eqref{eq:eg2phys}. The Borel singularity set is $\Gamma_w(z) = \{\chi_0(z),\chi_1(z),\chi_2(z)\}$. The curve may then be re-expressed in terms of the singulants as 
\begin{equation}
    P_z(w,\varphi) = 9(w-\chi_1(z))(w-\chi_2(z))\varphi^3 + \left( \frac{9 \sqrt{3}}{4}\right)^{2/3}\left(\chi_2(z) - \chi_1(z)\right)^{2/3}\varphi - 1 .
\end{equation}
It is then straightforward to read off the physical singularity sets from the germ expansions of this curve as:
\begin{equation}
    \Gamma_z^{(0)} = \{1\},\quad \Gamma_z^{(1)} = \{0\}, \quad \Gamma_z^{(2)} = \{0\}.
\end{equation}
We see that the point in $\Gamma_z^{(0)}$ is a $(0,1)$ turning point and the points in both $\Gamma^{(1),(2)}_z$ are $(1,2)$ turning points. We find, in addition, a virtual turning point at $z=1/4$ where $\pi \chi_2(z) = \pi \chi_0(z)$ but in fact $\chi_0(z)$ and $\chi_2(z)$ lie on different sheets of $\pi: \Sigma_z \to \mathbb{C}_w$. We note that although the singulants $\chi_i(z)$ are multi-valued functions of $z$, the curve $P_z(w,\varphi)$ is a single valued function on $\mathbb{C}_z$ (with meromorphic coefficients). The singularity graph is illustrated in figure \ref{fig:graphexample} and a sketch of the Riemann surface structure as $z$ varies is shown in figure \ref{fig:germexample}. In the following section we will discuss the Stokes phenomena associated to this curve.

\end{example}

\begin{example}
We conclude this section with a remark on a well-known example, the Pearcey equation \cite{howls2004higher}. The algebraic curve
\begin{equation}\label{eq:Pearcey-curve}
    P_z(\varphi,w) = (64w^3 + 32 w^2 -4w - 2z^2 - 72 w z^2 + 27z^4)\varphi^4 + (18z^2 + 8w +2)\varphi^2 + 8z \varphi + 1 = 0
\end{equation}
arises as the Borel transform of the formal solution to the Pearcey differential equation
\begin{equation}
    \epsilon^3 \frac{d^3 y}{dz^3} + \epsilon \frac{dy}{dz} + z y = 0.
\end{equation}
We note that the leading coefficient
\begin{equation}
    a_0(w,z) := (64w^3 + 32 w^2 -4w - 2z^2 - 72 w z^2 + 27z^4) = 64(w-\chi_1(z))(w-\chi_2(z))(w-\chi_3(z))
\end{equation}
has roots that set the discriminant to zero, namely $\Delta_P(\chi_i(z)) = 0$ for $i=1,2,3$. It is straightforward to verify that these are order three branch points and we describe the set $\Gamma_w \subset \Sigma_z$. Further studying the discriminant in $w$ for the polynoimal $a_0(w,z)$ we find that
\begin{equation}
    \Delta_{a_0} = 4096 \left(19683 z^8+8748 z^6+1296 z^4+64 z^2\right)
\end{equation}
Setting this discriminant to zero, we see the repeated non-zero solution
\begin{equation}
    z = \pm \frac{2i}{3 \sqrt{3}} ,
\end{equation}
thus describing the turning points where $\chi_i(z)$ coalesce and therefore the sets $\Gamma^{(\chi_i)}_z$.
We further verify that $\chi_i(z)$ defined as the solutions to $\Delta_P=0$ satisfy the characteristic equation of Pearcey:
\begin{equation}
    (\chi_i'(z))^3 + (\chi_i'(z)) + z = 0,\quad i=1,2,3,
\end{equation}
as expected.
\end{example}

\section{Stokes phenomenon}\label{sec:stokes}

In this section we discuss how to associate a stokes graph $\mathcal{G}$ to the algebraic curves $\Sigma_z$ discussed in the previous section. We then discuss a feature unique to parametric resurgence firstly developed by Howls \textit{et al.} and Berk \textit{et. al.} \cite{howls2004higher,chapman2007shock,berk1982new}, the higher-order Stokes phenomena. We elucidate this effect in our geometric setting.

\subsection{The inverse Borel transform}
\begin{figure}
    \centering
    \includegraphics[scale=0.2]{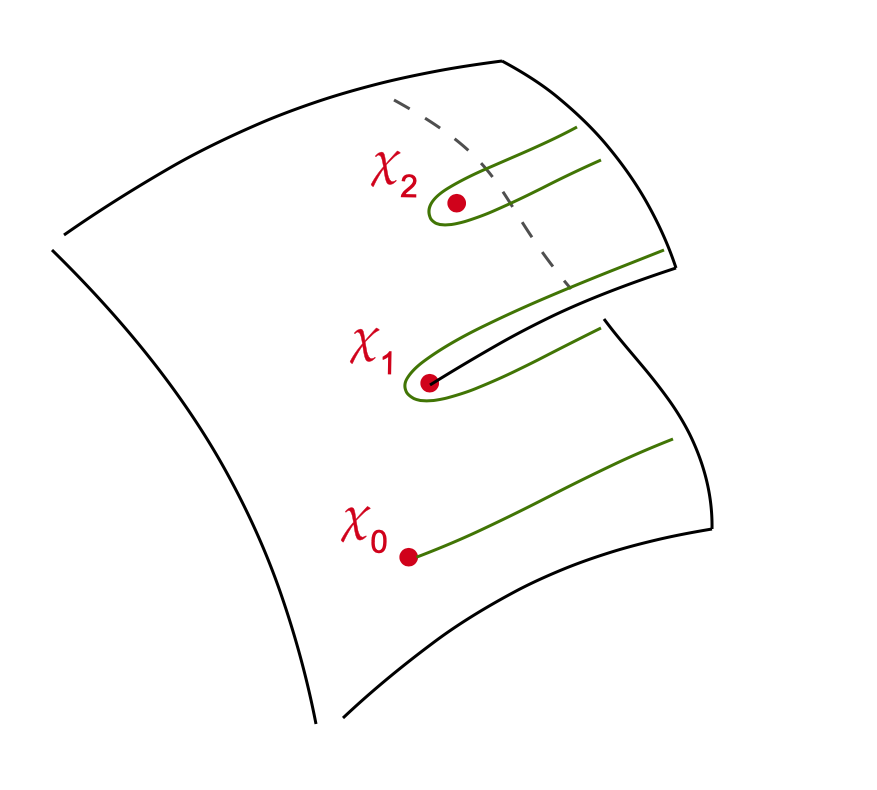}
    \caption{An illustration of $\Sigma_z$ for our running example. $\chi_0 = 0$ is also highlighted here since later we will consider this curve as a solution to an inhomogeneous problem where this point is distinguished.}
    \label{fig:Hankeleg}
\end{figure}

Let $\Sigma_z$ be a parametric algebraic curve of the type discussed in the previous section. We consider examples where $\Gamma_{w}(z)$ is a finite set and we can thus label singularities by $\chi_i$ with $i=1,\ldots, d$. Associated to each $\chi_i$ we may consider an oriented Hankel contour $\mathcal{H}_{\chi_i}$. We fix a particular direction of Hankel contour corresponding to a choice of $\epsilon > 0$. This setup of this section is illustrated in figure \ref{fig:Hankeleg}. Note that altering the phase of $\epsilon$ will lead to a mutated Stokes graph $\mathcal{G}_{\epsilon}$.

Consider the inverse Borel transform of $\Sigma_z$. This inverse Borel transform depends both on the curve and a contour choice
\begin{equation}
    \mathcal{H}' = \bigcup_{i=1}^d \sigma_{i}\mathcal{H}'_{\chi_i} \,.
\end{equation}
The primed Hankel contours denote half a Hankel contour to infinity. In the case that the Borel transform arises from a formal solution to a differential equation, the undetermined constants $\sigma_{i}$ will be related to a choice of initial/boundary data in the physical problem.

We may then write the inverse Borel transform of $(\Sigma_z,\mathcal{H})$ as an integral over appropriate analytically continued local germs of $\Sigma_z$ as 
\begin{equation}\label{eq:inverseHankelintegral}
    y(z,\epsilon) := \int_{\mathcal{H}'} dw \, e^{-w/\epsilon} \varphi_B(w;z) \,.
\end{equation}
Around each singularity $\chi_i$, recall that we have a local (singular) germ expansion of $\Sigma_z$ given by \eqref{eq:Borel-germ-at-chi-i} for each $\chi_i$:
\begin{equation}\label{eq:Borel-germ-exp}
    \Phi^{(i)}(w,z) = (w-\chi(z))^{-\alpha_i}\left( \Phi_0^{(i)}(z) + (w-\chi(z))\Phi_1^{(i)}(z) + \ldots\right) + \text{Reg.}
\end{equation}
We may then expand the integral \eqref{eq:inverseHankelintegral} asymptotically for small $\epsilon$ around each $\mathcal{H}_{\chi_i}$ to define a formal (finite) transseries
\begin{equation}\label{eq:egtransseries}
\begin{split}
    y(z,\epsilon;\sigma_1,\sigma_2,\ldots,\sigma_d) &= \sum_{i=1}^d \sigma_i \,\frac{e^{-\chi_i(z)/\epsilon}}{\epsilon^{\alpha_i-1}}\left(y_0^{(i)}(z) + \epsilon \,y_1^{(i)}(z) + \ldots \right) \\
    &= \sum_{i=1}^d \sigma_{i} e^{-\chi_i(z)/\epsilon}\,\epsilon^{1-\alpha_i}\,\sum_{k\ge0}\,\epsilon^k\,\Gamma(k+1-\alpha_i)\,\Phi_k^{(i)}(z), 
\end{split}
\end{equation}
as $\epsilon \to 0$. This is a transseries of the kind introduced in \eqref{eq:transs-sol-nth}. In this work, we will also consider formal transseries/Borel transformations associated to inhomogeneous equations. In this case we consider a marked Riemann surface $\Sigma_z$ with a distinguished point $p \in \Sigma_z$ (typically a pre-image of the origin $p \in \pi^{-1}$) and associate a contour $l = (0,\infty)$. The transseries then gains an additional $\chi_0(z) = 0$ component with $\alpha_0 = 0$ and we include $\chi_0$ in the set $\Gamma_w(z)$.

\paragraph{Stokes phenomena.}
The Stokes phenomena is now determined by the movement of the $\chi(z)$ in $\Gamma_w(z)$ as $z$ varies in $\mathbb{C}_z$. The coefficients $\sigma_{i}$ in \eqref{eq:egtransseries} undergo jumps across loci\footnote{Recall that we assume $\Sigma_z$ is single valued in $\mathbb{C}_z$ so we may unambiguously consider lines in $\mathbb{C}_z$ rather than on a cover.} $\ell_{i>j} \subset \mathbb{C}_z$ associated to pairs of singularities in $\Gamma_w(z)$ where the singularity for $\chi_j$ enters the Hankel contour $\mathcal{H}_i$ associated to $\chi_i$. Along this locus where the singularity $\chi_j$ intersects the contour $\mathcal{H}_{\chi_i}$ we say $i$ switches on $j$ and consequently we have a Stokes automorphism  $\mathfrak{S}_{i>j}$ acting on the transseries \eqref{eq:egtransseries} by
\begin{equation}
    \sigma_{j} \to \sigma_{j} + S_{ij}\sigma_i.
\end{equation}
This is directly linked to the analytic continuation of the germ $\Phi_i(w,z)$ to point $w=\chi_j$:
\begin{equation}
    \label{eq:borel-an-cont-stokes}
    \left.\Phi_i(w,z)\right|_{w=\chi_j}=-\frac{S_{ij}}{\nu_j}\,\Phi_j(w,z)+\cdots\,,
\end{equation}
where the Borel germs $\Phi_i(w,z)$ are given as expansions at the respective singularities $w=\chi_i(z)$ as in \eqref{eq:Borel-germ-exp}, $\nu_j$ is the discontinuity associated to the type of branch cut singularity in $\Phi_j$ (such as $\nu=2$ for a square root branch cut or $\nu=2\pi\mathrm{i}$ for a pole or log cut), and $S_{ij}$ is the associated Stokes constant. The collection of all Stokes lines $\bigcup_{(i,j)} \ell_{i>j}$ is called a Stokes graph $\mathcal{G}_{\epsilon}\subset \mathbb{C}_z$.

On the other hand, the na\"{i}ve Stokes line condition of the formal transseries expansion \eqref{eq:egtransseries} correspond instead to relations in $\hat{\Gamma}_w(z)$:
\begin{equation}
    \hat{\ell}_{i>j}: \, \Re \pi \circ \chi_i(z) > \Re \pi \circ \chi_j(z), \quad \Im \pi \circ \chi_i(z) = \Im \pi \circ \chi_j(z) .
\end{equation}
These lines then emanate from the previously discussed turning points $z_{ij}$ between $\chi_i$ and $\chi_j$. The collection of all $\hat{\ell}_{i>j}$ form a \textit{na\"{i}ve Stokes graph} on $\mathbb{C}_z$.

Since $(\Sigma_z,\pi)$ has a multi-sheeted structure in general the true Stokes line $\ell_{i>j}$ will be truncated compared to $\hat{\ell}_{i>j}$ so that $\ell_{i>j} \subset \hat{\ell}_{i>j}$. This truncation phenomenon is referred to as higher-order Stokes phenomena. We return to this feature in more detail shortly.

\subsection{Coefficient transseries}

We now turn to discuss a distinct but related transseries structure that we may also naturally associate to a curve $\Sigma_z$. Recall first that around each singular point $\chi$ we have a germ expansion \eqref{eq:Borel-germ-exp}.
One may then consider a parametric transseries associated to the large $n$ expansion of the coefficients themselves $\Phi^{(i)}_n(z)$. Darboux's theorem (see \textit{e.g.} \cite{dingle1973asymptotic}) tells us that starting from a non-singular expansion at $w=\chi_i$ ($\alpha_i=0$), the large $n$ behaviour of the expansion coefficients is generically of the form
\begin{equation}\label{eq:Dexpansion}
\Phi_n^{(i)}(z) \sim \sum_{\chi_{j} }\frac{(-1)^{-\alpha_j}\Gamma(n+\alpha_j)}{\chi_{ij}(z)^{n+\alpha_j}\Gamma(n+1)} \left( \frac{\Phi_0^{(j)}(z)}{\Gamma(\alpha_j)} -
    \frac{1}{(n+\alpha_j-1)}\frac{\chi_{ij} \,\Phi_1^{(j)}(z)}{\Gamma(\alpha_j-1)}	\\
		+ \ldots \right) \,,
\end{equation} 
where $\chi_{ij}=\chi_j-\chi_i$ and $\chi_j$ are the neighbouring saddle points to $\chi_i$. The exponential weights/actions of this transseries may be written as $e^{(n+\alpha)\log \chi}$, suggestive of a relationship between the original Borel transform $\varphi_B(w,z)$ with respect to $\epsilon$ and the Borel transform of the large $n$ expansion of $\Phi_n^{(i)}$ up to the exponential conformal map. To investigate this relationship we will, for concreteness, focus on our running example \ref{eg:runninggerms}.

\paragraph{Coefficient Borel transform.}
In our running example, we consider $\chi_0(z) = 0$ as a distinguished point giving the Borel transform of the particular integral of the inhomogeneous equation. Let us then study the large $n$ expansion of the coefficients of the germ about the origin $\{\Phi^{(0)}_n(z)\}_{n\ge 0}$. 

\begin{figure}
    \centering
    \includegraphics[scale=0.2]{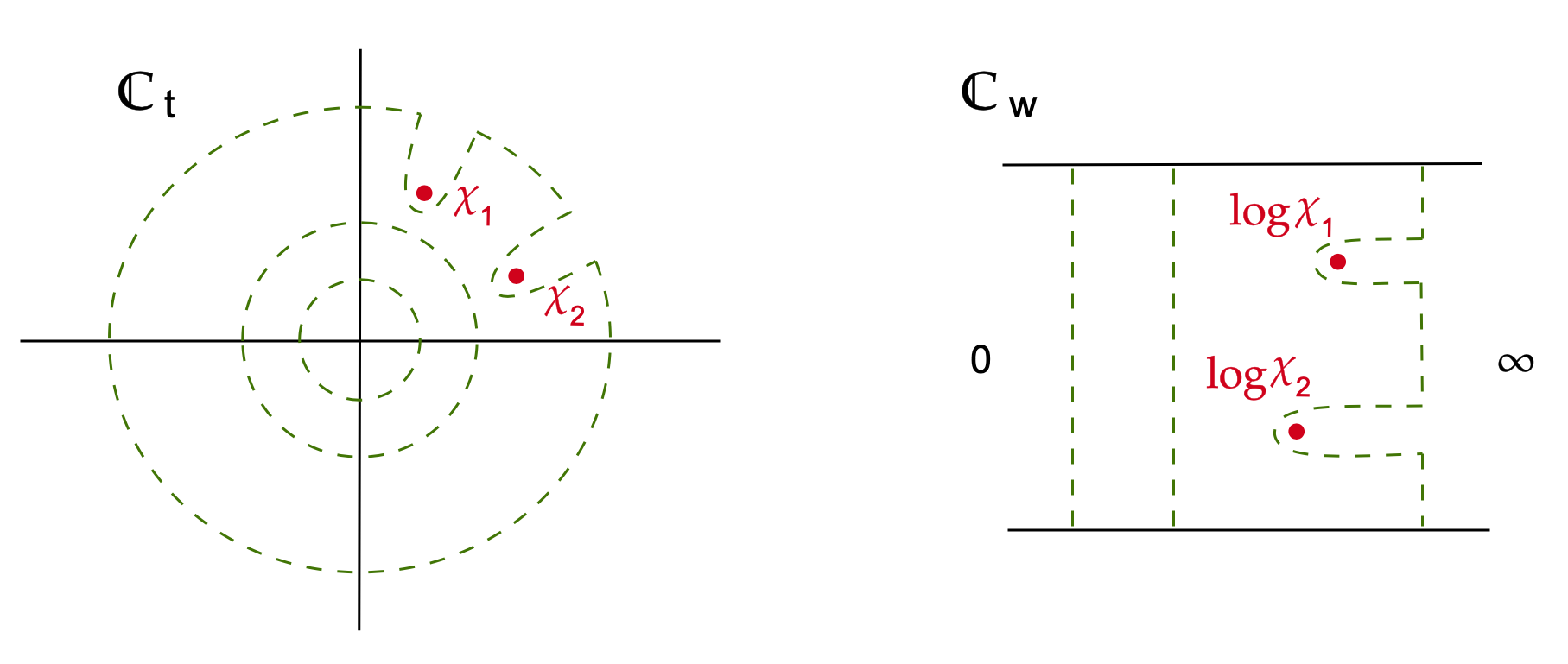}
    \caption{The exponential conformal map showing relation between Stokes lines in original and conformal planes. The dashed lines and the different integrations contours. Stokes lines emanating from $\log \chi$ now appear. }
    \label{fig:coefficients}
\end{figure}

We first write a Cauchy integral formula for the coefficients 
\begin{equation}
    \Phi_n^{(0)}(z) = \frac{1}{2 \pi i} \oint_{S^1} \frac{dt}{t^{n+1}} \varphi_B(t;z) \,,
\end{equation}
where $\varphi_B(t;z)$ denotes the analytic continuation of the Borel germ from our distinguished origin and the radius of the circle is smaller than the distance to the nearest singularity. We recover an inverse Borel transform integral by considering a conformal change of variable $t=e^w$ so that
\begin{equation}\label{eg:coefficientsaddle}
    \Phi_n^{(0)}(z) = \frac{1}{2 \pi i} \oint_{C} dw \,e^{-w n}\varphi_B(e^w;z) \,.
\end{equation}
In our example, the integrand has potential singularities at $w = \log \chi_1$ and $w = \log \chi_2$ with associated steepest descent contours as illustrated in figure \ref{fig:coefficients}. 

\begin{remark}We note here that during the preparation of the manuscript a related analysis of the conformal relationship between the $1/n$ transseries and the $\epsilon$ transseries in the non-parametric setting appeared in \cite{Marinissen:2023ttp}. The construction is also discussed in the work \cite{crew2024resurgent}.
\end{remark}

\paragraph{Stokes phenomenon.}
With the change of variables illustrated in figure \ref{fig:coefficients} we may now consider deforming the contour to perform a steepest descent analysis for large $n$. From this figure we see that the above integral undergoes Stokes phenomenon across the Stokes line of the large $n$ variable:
\begin{equation}
    \Re \log \chi_2(z) > \Re \log \chi_1(z) \,, \quad \Im \log \chi_2(z) = \Im \log \chi_1(z) \,,
\end{equation}
where the steepest descent contours change. The locus is equivalently expressed as
\begin{equation}
    \arg \chi_1(z) = \arg \chi_2(z) \,, \quad |\chi_1(z)|<|\chi_2(z)| \,.
\end{equation}
and concerns the arrangement of three points $\chi_0=0$, $\chi_1$ and $\chi_2$ as it describes the presence (or not) of the $\log \chi_2$ saddle point in the large $n$ transseries of the germ coefficients at $\chi_0$. We thus label the Stokes line as $h_{012}$ --- it is illustrated in figure \ref{fig:stokesgrapheg}. We will see momentarily that whilst $h_{012}$ is an ordinary Stokes line of the $1/n$ expansion, it is in fact a \textit{higher-order Stokes line} of the $\epsilon$ asymptotic expansion.

We may then evaluate the leading saddle point asymptotics of \eqref{eg:coefficientsaddle} as an illustrative example and we see that we recover the $1/n$ formal transseries of the form \eqref{eq:Dexpansion}, explicitly
\begin{equation}
    \Phi_n^{(0)}(z) \sim \frac{\Gamma(n+\frac{1}{2})}{\Gamma(n+1)\Gamma(\frac{1}{2})}\frac{(-1)^{-\frac{1}{2}}}{\chi_1(z)^{n+\frac{1}{2}}}\left(\frac{(-2)^{-1/2}}{(1-z)} + \frac{\chi_1(z)}{2\left(n-\frac{1}{2}\right)}\frac{(-2)^{1/2}}{(1-z)^3}+\ldots \right) - \frac{\tau_2}{\chi_2(z)^{n+1}},
\end{equation}
 where higher coefficients can be obtained from the germ expansions of section \ref{eg:runninggerms}.

The parameter $\tau_{2}$ is translated along the locus $h_{012}$ (the higher-order Stokes line). It is clear that the higher-order Stokes phenomena  affects the ordinary Stokes phenomena since the presence (or not) of such a term in the late order behaviour determines whether (or not) the Stokes line for $\chi_2$ is active. Geometrically the locus $h_{012}$ has a clear interpretation since this determines when $\chi_2$ is visible to $\chi_0$ on the Riemann surface and it thus determines where the na\"{i}ve Stokes line $\hat{\ell}_{02}$ is truncated to $\ell_{02}$.

\subsection{Higher-order Stokes phenomenon}
Generally, we may consider any triple of singular points $\chi_1$,$\chi_2$ and $\chi_3$ in $\Gamma_w(z)$. After the conformal map centred on $\chi_1$, it is straightforward to see that the $1/n$ transseries arising from the Cauchy integral for the $\Phi^{(\chi_1)}_n$ coefficients may have a Stokes line along the locus
\begin{equation}
    h_{123}: \quad \Im \log(\chi_2-\chi_1) = \Im \log(\chi_3-\chi_1), \quad \Re \log(\chi_2-\chi_1) < \Re \log(\chi_3-\chi_1)
\end{equation}
or equivalently,
\begin{equation}
    h_{123}: \quad \arg (\chi_2-\chi_1) = \arg (\chi_3 - \chi_1), \quad |\chi_2-\chi_1| < |\chi_3 - \chi_1|\,.
\end{equation}
In other words, the projections $\pi \chi_1$, $\pi \chi_2$ and $\pi \chi_3$ are collinear in $\mathbb{C}_w$. Along this locus, the contribution of $\chi_3$ to the $1/n$ transseries undergoes Stokes phenomena. With respect to the ordinary  Stokes phenomena in $\epsilon$ then the line $h_{123}$ describes the ability of $\chi_3$ to be turned on by $\chi_1$. Geometrically it is the condition for $\chi_3$ to be visible to $\chi_1$ on $\Sigma_z$. It is in this sense that $h_{123}$ is a higher-order Stokes line -- the higher-order Stokes line describes the possible truncation $\ell_{13} \subset \hat{\ell}_{13}$ caused by $\chi_2$.

In terms of the Stokes graph on $\mathbb{C}_z$ it is clear that a higher-order Stokes line $h_{123}$ emanates from points in $\mathbb{C}_z$ where $\chi_2$ and $\chi_3$ coincide (Stokes crossing point). In general, a na\"{i}ve $\hat{\ell}_{13}$ Stokes line emanating from a virtual turning point between $\chi_1$ and $\chi_3$ is inactive until the point where $\chi_2$ and $\chi_3$ cross.

\begin{figure}
    \centering
    \includegraphics[width=6.7cm]{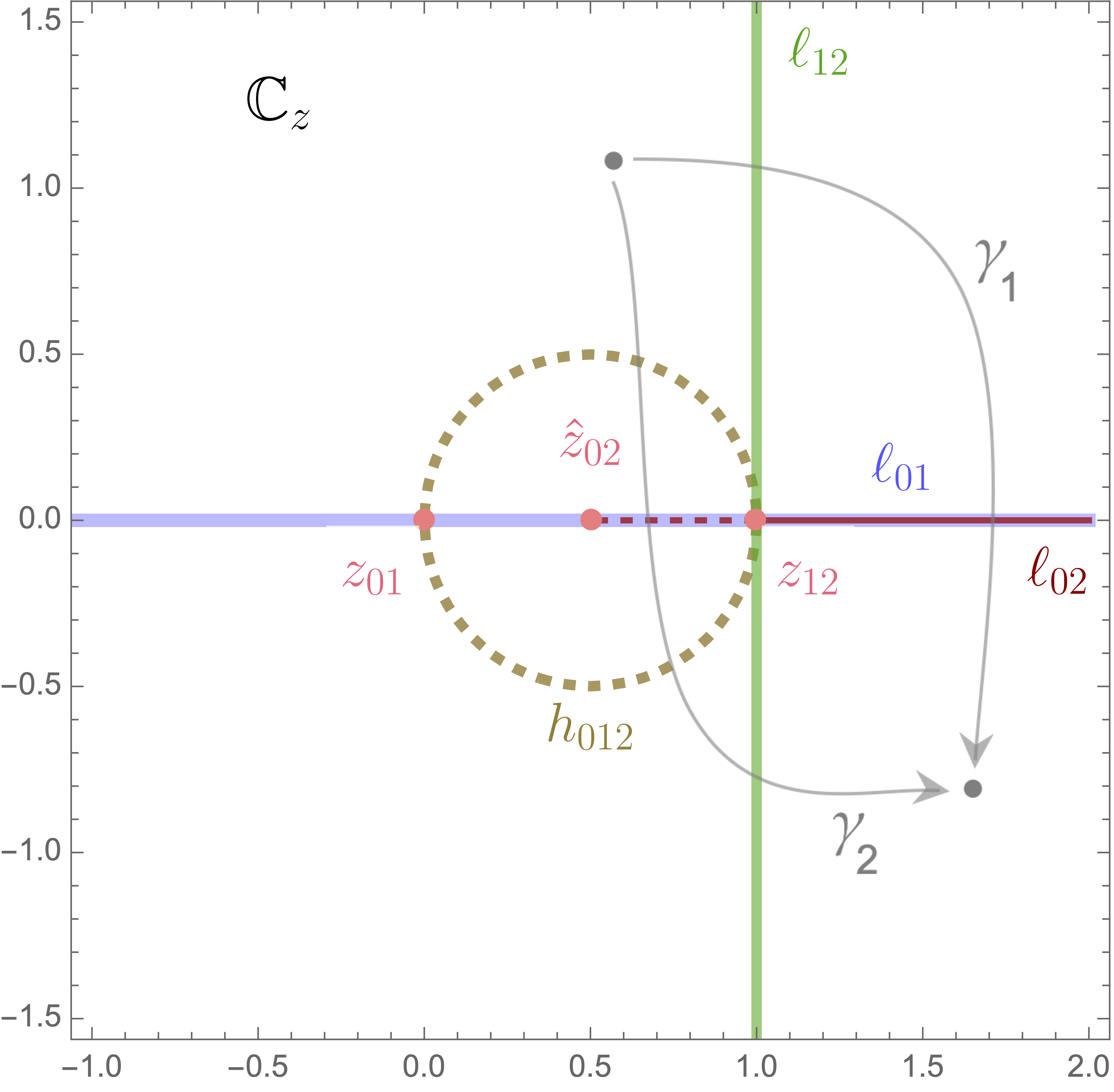}
    \caption{This is the Stokes graph in $\mathbb{C}_z$ for our running example \eqref{eq:curve-main-example}. In the terminology of section \ref{sec:algcurves}, $z=0$ and $z=1$ are genuine $(0,1)$ and $(1,2)$ turning points respectively whereas $\hat{z}=1/2$ is a virtual turning point. The Stokes line $\ell_{01}$ is shown in light blue, while $\ell_{12}$ is shown in green. The na\"{i}ve Stokes line $\hat{\ell}_{02}$ (red) is truncated along the higher-order stokes line (dashed brown) $h_{012}$. Finally two illustrative paths $\gamma_1$ and $\gamma_2$ of analytic continuation are shown, the latter undergoing higher-order Stokes phenomena. }
    \label{fig:stokesgrapheg}
\end{figure}

\begin{example}
We can now describe the full Stokes graph for our running example \eqref{eq:curve-main-example}. Recall that the singulants in $\hat{\Gamma}_w(z)$ are given by
\begin{equation}
    \chi_0(z) = 0, \quad \chi_1(z) = \frac{z^2}{2}, \quad \chi_2(z) = z-1/2.
\end{equation}
We first consider the na\"{i}ve Stokes lines
\begin{equation}
\begin{split}
    \hat{\ell}_{01}:& \quad \Re \frac{z^2}{2}>0, \quad \Im \frac{z^2}{2} = 0 \\
    \hat{\ell}_{02}:& \quad \Re z > 1/2, \quad \Im z = 0 \\
    \hat{\ell}_{12}:& \quad \Re \frac{z^2}{2} < \Re (z-1/2), \quad \Im \frac{z^2}{2} = \Im z \\
\end{split}
\end{equation}
with associated turning points discussed in \eqref{eq:egturningpoints}. We have, in addition, the higher-order Stokes line
\begin{equation}
    h_{012}: |z-1/2| = |z| \,.
\end{equation}
This information is collected together in figure \ref{fig:stokesgrapheg}, where one clearly sees that the na\"{i}ve Stokes graph is truncated as expected.

The Stokes phenomenon for this example can be described by the analytic continuation of the germs $\Phi_i(w,z)$ with $i=0,1,2$ which were determined in the previous section, in the form given by \eqref{eq:borel-an-cont-stokes}:
\begin{eqnarray}
     \left.\Phi_0(w,z)\right|_{w=\chi_1} \hspace{15pt} & = & -\frac{S_{01}}{2}\,\Phi_1(w,z)+\cdots\,;\\
      \left.\Phi_0(w,z)\right|_{w=\chi_2} \hspace{15pt} & = & -\frac{S_{02}}{2\pi\mathrm{i}}\,\Phi_2(w,z)+\cdots\,;\\
       \left.\Phi_1(w,z)\right|_{w=\chi_2-\chi_1} & =& -\frac{S_{12}}{2\pi\mathrm{i}}\,\Phi_2(w,z)+\cdots\,.
\end{eqnarray}
The germs above are defined by their expansions at the respective singular points, and the Stokes constants can then be read directly from a large order analysis of the coefficients (or from the respective algebraic curve in our examples). These Stokes constants are given by:
\begin{eqnarray}
    S_{01} &=& 2\,;\label{eq:piecewise-S01}\\
    S_{12} &=& -\pi\mathrm{i}\,;\label{eq:piecewise-S12}\\
    S_{02} &=& \left\{\begin{matrix}
-2\pi\mathrm{i},\,&z\in \mathbb{C}_z\backslash R_{\mathrm{h.o.s}};\\  
0,\quad& z\in R_{\mathrm{h.o.s}}\,.\hspace{17pt}
    \end{matrix}\right.\label{eq:piecewise-S02}
\end{eqnarray}
The region $R_{\mathrm{h.o.s.}}$ is the region enclosed by the higher-order Stokes line $h_{012}$ in figure \ref{fig:stokesgrapheg}. The discrete jump in the Stokes constant $S_{02}$ here reflects the higher-order Stokes phenomenon introduced above and is discussed in more detail section \ref{subsec:automorphisms} below.
\end{example}

\begin{figure}
    \centering
    \includegraphics[width=6.6cm]{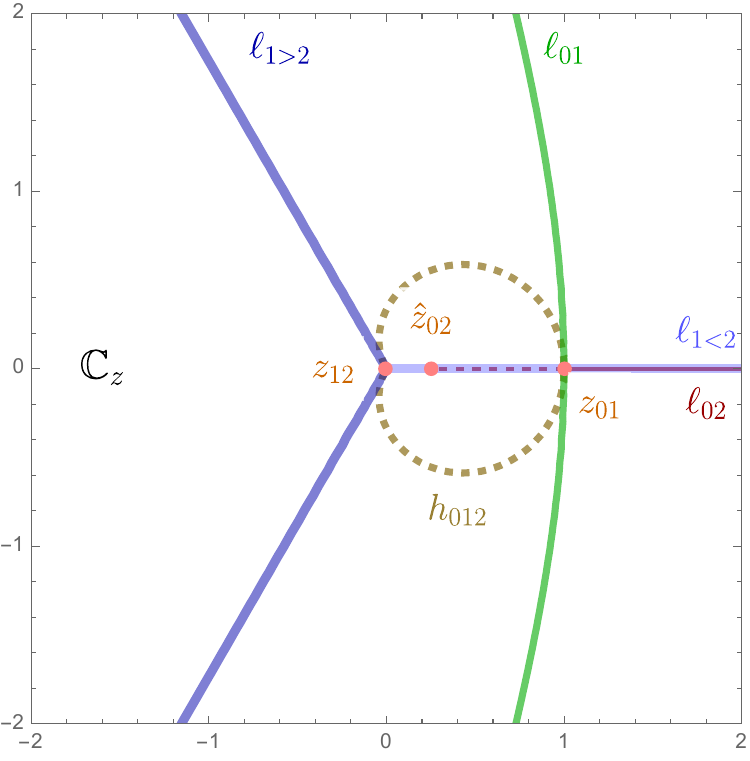}
    \caption{This is the Stokes graph in $\mathbb{C}_z$ for the second example \eqref{eq:eg2curve}. As discussed in section \ref{sec:algcurves}, $z=0$ and $z=1$ are $(1,2)$ and $(0,1)$ turning points respectively, and $\hat{z}=1/4$ is a virtual turning point. Stokes lines $\ell_{1>2}$ and $\ell_{1<2}$ are shown in dark and light blue respectively, while $\ell_{01}$ is shown in green. The cardioidal higher-order Stokes line is shown in dashed brown, truncating the na\"{i}ve Stokes line $\hat{\ell}_{02}$ (red).}
    \label{fig:stokesgrapheg2}
\end{figure}

\begin{example}
We also illustrate the ideas of this section for our second main example \ref{eg:second}. Recall that the singulants in $\hat{\Gamma}_w(z)$ are given by
\begin{equation}
    \chi_0(z) = 0, \quad \chi_1(z) = \frac{1}{3}(-1+3z - 2 z^{3/2}), \quad \chi_2(z) = \frac{1}{3}(-1+3z + 2 z^{3/2}).
\end{equation}    
The na\"{i}ve Stokes lines
\begin{equation}
\begin{split}
    \hat{\ell}_{01}:& \quad \Re \chi_0(z) < \Re \chi_1(z), \quad \Im \chi_0(z) = \Im \chi_1(z), \\
    \hat{\ell}_{02}:& \quad \Re \chi_0(z) < \Re \chi_2(z), \quad \Im \chi_0(z) = \Im \chi_2(z), \\
    \hat{\ell}_{12}:& \quad \Re \chi_1(z) < \Re \chi_2(z), \quad \Im \chi_1(z) = \Im \chi_2(z), \\
\end{split}
\end{equation}
emanate from the turning points discussed earlier and are illustrated in figure \ref{fig:stokesgrapheg2}. We have again a higher-order Stokes line whereby $\chi_1$ obstructs the $0>1$ Stokes line. This is the locus
\begin{equation}
    h_{012}: \, \arg \chi_2(z) = \arg \chi_1(z) \,.
\end{equation}
The higher-order Stokes line is determined numerically and the complete Stokes graph, together with the turning points discussed previously in example \ref{eg:second}, is given in figure \ref{fig:stokesgrapheg2}.
\end{example}

\paragraph{Remarks.}\label{rmk:hoho} The present section may be summarised as follows. We associate a formal transseries to a parametric Riemann surface $\Sigma_z$ with singularity set $\Gamma_w(z)$, with a collection of constants (transseries parameters) $\sigma_i$ for each $\chi_i\in\Gamma_w(z)$, which undergo Stokes automorphisms. For each singular germ at $\chi$ there is another $1/n$ transseries whose Stokes graph is essentially determined by a conformal map, $\varphi_B(e^w,z)$, leading to higher-order Stokes lines that arise as logarithms of the usual Dingle dominance/anti-dominance conditions. It would be interesting to investigate the iteration of this procedure whereby the germ expansion of $\varphi_B(e^w,z)$ at singularities $w=\log \chi$ may be further expanded as a $1/n$ transseries giving a Stokes structure determined by $\varphi_B(e^{e^w},z)$. In this way, for problems with at least four singularities we expect to find higher higher-order Stokes lines that switch off the activity of higher-order Stokes lines. According to the iterated conformal map, these would be located at the following locus in $\mathbb{C}_z$:
\begin{equation}
\begin{split}
    \Im \log \left(\log(\chi_4-\chi_1)-\log(\chi_2-\chi_1
)\right)) &= \Im \log \left(\log(\chi_3-\chi_1)-\log(\chi_2-\chi_1)\right), \\
    \Re \log \left(\log(\chi_4-\chi_1)-\log(\chi_2-\chi_1
)\right) & > \Re \log \left(\log(\chi_3-\chi_1)-\log(\chi_2-\chi_1)\right). \\
\end{split}
\end{equation}
One may encounter such phenomena for example by adding an inhomogeneous term to the 3rd order Pearcey equation corresponding to the germ expansion of the algebraic curve solution \eqref{eq:Pearcey-curve} at $w=0$.

\subsection{Stokes automorphisms}\label{subsec:automorphisms}
We now discuss Stokes automorphisms. Let us consider a curve (open or closed) $\gamma \subset \mathbb{C}_z$ with an associated Stokes automorphism $\mathfrak{S}_{\gamma}$. In general, the automorphism acts on transseries \eqref{eq:egtransseries} by a finite translation of the transseries parameters\footnote{This is the analogue of analytic continuation in the argument of the variable in equational resurgence.}
\begin{equation}
    \mathfrak{S}_{\gamma}: y(z,\epsilon;\sigma_0,\sigma_1,\sigma_2) \to y(z,\epsilon;\sigma_0',\sigma_1',\sigma_2') 
\end{equation}
To determine this action, the total automorphism across $\gamma$ may be decomposed into generators $\mathfrak{S}_{ij}$ \textit{i.e.} Stokes automorphisms associated to a curve $\gamma_{ij}$ that crosses a single Stokes line $\ell_{ij}$. Our running example \ref{eg:runningexample} has the following generators
\begin{equation}
\begin{split}
    &\mathfrak{S}_{01}: \sigma_1 \to \sigma_1 + S_{01}\,\sigma_0, \\
    &\mathfrak{S}_{12}: \sigma_2 \to \sigma_2 + S_{12}\,\sigma_1, \\
    &\mathfrak{S}_{02}: \sigma_2 \to \sigma_2 + S_{02}\,\sigma_0.
\end{split}    
\end{equation}
where the Stokes constants are defined in equations \eqref{eq:piecewise-S01}, \eqref{eq:piecewise-S12} and \eqref{eq:piecewise-S02}.

In comparison with equational/constant resurgence the Stokes automorphism group has more structure in the co-equational/parametric case, namely the Stokes constants may be only piecewise constant, as we shall now explain. 

Let us first consider na\"{i}ve Stokes automorphism generators $\hat{\mathfrak{S}}_{ij}$ associated to na\"{i}ve Stokes lines $\hat{\ell}_{ij}$. In the parametric setting we have another set of Stokes automorphisms $\mathfrak{T}_h$ associated to higher-order Stokes lines $h \subset \mathbb{C}_z$. In contrast with the Stokes automorphisms $\mathfrak{S}$, the $\mathfrak{T}_h$ automorphisms act on the $1/n$ transseries associated to the germ expansion of each Borel singularity. In our running example, only the $1/n$ transseries associated to the germ at $\mathbf{0} \in \Sigma_z$ undergoes non-trivial $\mathfrak{T}_h$ Stokes automorphisms. Such formal $1/n$ transseries may be expressed in a general descending factorial form as
\begin{equation}\label{eq:Stokes-aut-running}
    g^{\mathbf{0}}(z,n;\tau_1,\tau_2,\ldots, \tau_{d-1}) = \sum_{i=1}^{d-1} \tau_i^{\mathbf{0}} \frac{\Gamma(n+\alpha_i)}{\Gamma(n+1)}\left( g^{\mathbf{0}}_0(z) + \frac{1}{n+\alpha_i-1}g^{\mathbf{0}}_1(z) + \ldots
    \right),
\end{equation}
where the sum is taken over all other singularities in $\Gamma_w \subset \Sigma_z$ and the visibility, or not, of a singularity $\chi_i$ on the Riemann sheet of $\mathbf{0}$ is determined by whether the value of the transseries parameter $\tau_i^{\mathbf{0}}$ is one or zero. We note here that, unlike the $\epsilon$ transseries parameters, the $\tau_i$ parameters are not free parameters but are fixed according to the contour specified by the Cauchy integral in \eqref{eg:coefficientsaddle}. In the present context this amounts to a choice of branch for $\mathbf{0}$.

Returning to our running example, the Stokes automorphism $\mathfrak{T}_h$ now acts as:\footnote{The appearance of $\text{mod}\, 2$ here is due to the exponential conformal map relation between $\epsilon$ and $1/n$ transseries, thus transseries parameters of $1/n$ transseries appear naturally in exponents $e^{2\pi i}$.}
\begin{equation}
    \mathfrak{T}_h: \tau_2^{\mathbf{0}} \to \tau_2^{\mathbf{0}} + 1 \quad \text{mod}\,2
\end{equation}
and trivially on the other transseries parameters. The automorphism $\mathfrak{T}_h$ is thus responsible for the Stokes phenomenon of the Stokes constants of the $\epsilon$-transseries themselves, leading to the piecewise constant nature of $S_{02}$ on $\mathbb{C}_z$ seen previously in equation \eqref{eq:piecewise-S02}.

There is a geometric perspective on the piecewise constant nature of Stokes constants whereby the projection $\pi: \Sigma_z \to \mathbb{C}_w$ induces a quotient on the Stokes automorphism group. Considering na\"{i}ve Stokes lines $\hat{\ell}_{ij}$ and their corresponding Stokes automorphism generators $\hat{\mathfrak{S}}_{ij}$ (which satisfy the relation $\hat{\mathfrak{S}}_{ij} = \hat{\mathfrak{S}}_{ji}^{-1}$) corresponds to working with the projected singularities $\pi \chi(z) \subset \mathbb{C}_w$. We must then also include the action of higher order Stokes automorphisms $\mathfrak{T}_{ij;k}$ where the $\hat{\ell}_{ij}$ Stokes line is disrupted by the presence of a singularity $\chi_k$. This collection of operators, which we denote $\text{Aut}(\hat{\Gamma}_w)$, may be considered to act as translations on an extended flat transseries parameter space
\begin{equation}
(\sigma_1,\tau^{(1)}_2,\tau^{(1)}_3,\ldots,\tau^{(1)}_d;\sigma_2,\tau^{(2)}_1,\tau^{(2)}_3,\ldots,\tau^{2}_d;\ldots;\sigma_d,\tau^{(d)}_1,\tau^{(d)}_2,\ldots \tau^{(d)}_{d-1}),
\end{equation}
comprising of the parameters $\{\sigma_i\}$ of the $\epsilon$-transseries together with the parameters $\{\tau^{(l)}_j\}_{l \neq j}$ for each $1/n$ coefficient transseries corresponding to each singular germ. The pre-quotient automorphism group $\text{Aut}(\hat{\Gamma}_w)$ then acts as 
\begin{equation}
    \mathcal{T}_{ij;k}: \tau^{(i)}_j \to \tau^{(i)}_j + 1 \quad \text{mod}\,2,
\end{equation}
(and identically for all other constants $\tau_l^{(m)}$) for the higher-order automorphisms and
\begin{equation}
    \hat{\mathfrak{S}}_{ij}: \sigma_j \to \sigma_j + \hat{S}_{ij} \,\tau^{(i)}_j\,\sigma_i ,
\end{equation}
for the naive Stokes automorphisms, where $\hat{S}_{ij}$ is the non-zero value of the piecewise constant Stokes constant. Intuitively, $\chi_i$ may only switch on $\chi_j$ when $\chi_j$ is visible on the Riemann sheet associated to $\chi_i$--that is $\tau^{(i)}_j$ is non-zero. Taking the quotient determined by the relations satisfied by the generators, namely\footnote{The order of these relations requires one to pick an ordering of the sheets of $\pi:\Sigma_z \to \mathbb{C}_w$, this is realised in our example by a distinguished choice of $\mathbf{0}$.}
\begin{equation}
    \mathfrak{T}_{ij;k} \circ \hat{\mathfrak{S}}_{ij} = \mathfrak{S}_{ij}, \quad \hat{\mathfrak{S}}_{ij} \circ \mathfrak{T}_{ij;k} = \text{Id.}
\end{equation}
when crossing into a higher-order Stokes region recovers the Stokes automorphism group $\text{Aut}(\Sigma_z)$ acting now solely on the transseries parameters $(\sigma_1,\ldots,\sigma_n)$. 
This construction may be viewed as lifting to the Riemann surface $\Sigma_z$ at the cost of introducing piecewise constant (on $\mathbb{C}_z$) $\epsilon$-transseries Stokes constants defined by $S_{ij} := \hat{S}_{ij}\tau^{(i)}_j$.

The total Stokes automorphism across a curve $\gamma \subset \mathbb{C}_z$ may then be decomposed into generators according to the lines crossed by $\gamma$ in the Stokes graph $\mathcal{G}$,  giving a word built from generators $\hat{\mathfrak{S}}_{\hat{\ell}}$ and $\mathfrak{T}_h$. In the more general case, one should also include contributions from the monodromy group of the governing differential equation. We note that, passing to the quotient, one may construct an equivalence relation on paths in $\mathbb{C}_z$ corresponding to word equivalence in $\text{Aut}(\Sigma_z)$.

For example, let us consider the two paths $\gamma_1$ and $\gamma_2$ illustrated on figure \ref{fig:stokesgrapheg}. Since the curve $P_z$ \eqref{eq:curve-main-example} is single-valued on $\mathbb{C}_z$ and equation \eqref{eq:running-example-ODE} has no monodromy at finite $z$, we must have $\mathfrak{S}_{\gamma_1} \circ \mathfrak{S}_{\gamma_2^{-1}} = \text{Id}$. The first path $\gamma_1$ may be decomposed into na\"{i}ve automorphism generators as 
\begin{equation}
    \hat{\mathfrak{S}}_{\gamma_1} = \hat{\mathfrak{S}}_{02}\circ\hat{\mathfrak{S}}_{01}\circ \hat{\mathfrak{S}}_{12},
\end{equation}
and the second path $\gamma_2$ may be decomposed as
\begin{equation}
    \hat{\mathfrak{S}}_{\gamma_2} = \hat{\mathfrak{S}}_{12}\circ\mathfrak{T}_{012}\circ\hat{\mathfrak{S}}_{01}\circ\hat{\mathfrak{S}}_{02}\circ\mathfrak{T}_{012}
\end{equation}
The higher-order Stokes phenomena manifests in this algebraic structure as the following relations
\begin{equation}
\begin{split}
   \hat{\mathfrak{S}}_{02}\circ \mathfrak{T}_{012} = \text{Id}, \quad  \mathfrak{T}_{012} \circ \hat{\mathfrak{S}}_{02} = \mathfrak{S}_{02}.
\end{split}   
\end{equation}
which may then be applied to give the total Stokes automorphisms
\begin{equation}
    \mathfrak{S}_{\gamma_1} = \mathfrak{S}_{02}\circ\mathfrak{S}_{01}\circ \mathfrak{S}_{12},\quad
    \mathfrak{S}_{\gamma_2} = \mathfrak{S}_{12}\circ\mathfrak{S}_{01}.
\end{equation}
Finally, $\mathfrak{S}_{\gamma_1} \circ \mathfrak{S}_{\gamma_2^{-1}} = \text{Id}$ may be verified by using the value of the Stokes constants \eqref{eq:piecewise-S01}, \eqref{eq:piecewise-S12}, \eqref{eq:piecewise-S02} in the Stokes automorphisms \eqref{eq:Stokes-aut-running}:
\begin{eqnarray}
    \mathfrak{S}_{\gamma_1}y(z,\epsilon;\sigma_0,\sigma_1,\sigma_2) &=& y(z,\epsilon;\sigma_0,\sigma_1+S_{01}\sigma_0,\sigma_2+S_{12}\sigma_1+S_{02}\sigma_0)\,;\\
\mathfrak{S}_{\gamma_2}y(z,\epsilon;\sigma_0,\sigma_1,\sigma_2) &=& y(z,\epsilon;\sigma_0,\sigma_1+S_{01}\sigma_0,\sigma_2+S_{12}\sigma_1+S_{12}S_{01}\sigma_0)\,.
\end{eqnarray}
and the observation $S_{12}S_{01}=S_{02}=-2\pi\mathrm{i}$. Note that when a path crosses a Stokes half-line $\ell_{ij}$ (emanating from a turning point), it gives rise to an automorphism or its inverse depending on the direction of crossing.

\begin{remark}
It would be interesting to investigate the infinitesimal version of the above construction. In the co-equational resurgence context, the algebra of $z$-dependent Alien derivatives $\Delta_{\chi(z)}$ generate the automorphism group $\text{Aut}(\Sigma_z)$. Projecting the Riemann sheet structure one expects a commutative algebra of Alien derivatives $\hat{\Delta}_{\pi \chi}$ generating $\text{Aut}(\Gamma_w)$. Including the Alien derivative operators corresponding to the analytic continuations of $\Phi^{(i)}(e^{w},z)$ one expects to recover the Alien derivative algebra acting on $\Sigma_z$.
\end{remark}

\begin{remark}
Parallel to the remark on page \pageref{rmk:hoho}, one may also anticipate the possible inclusion of na\"{i}ve \textit{higher}-order automorphisms $\hat{h}_{ijk}$ whose activity is disrupted by a fourth singularity $\chi_l \in \Gamma_w$ across a higher higher-order Stokes line $h_{ijkl}$ giving rise to additional relations in the automorphism group. 

More precisely, consider a singularity $\chi_i$ and the associated $1/n$ transseries components with transseries parameters $\tau^{(i)}_j$ with $j = 1,\ldots,i-1,i+1,\ldots,d$. We may then consider a further transseries associated to each of these components with transseries parameters $\tau^{(i,j)}_k$ with $k$ not equal to $i$ or $j$ which may undergo a further Stokes automorphism
\begin{equation}
    \mathfrak{T}_{ijkl}: \tau^{(i,j)}_k \to \tau^{(i,j)}_k + 1 \quad \text{mod}\,2
\end{equation}
The pre-quotient automorphism $\hat{h}_{ijk}$ is then modified according to
\begin{equation}
    \hat{h}_{ijk}: \tau_j^{(i)} \to \tau_j^{(i)} + \tau^{(i,j)}_k.
\end{equation}
Consequently the Stokes constant $S_{ij}$ now obtains a more complicated piecewise dependence since the $1/n$ Stokes constants are now also piecewise constant (one or zero). We leave a more general investigation of this algebraic structure to future work.
\end{remark}

\section{Outer transseries analysis}
\label{sec:Outer}

In the previous section we analysed Borel transforms as solutions to Borel PDEs/singularly perturbed differential equations and whose analytic continuation satisfied an algebraic relation. In this section we turn to an analysis of the physical problem without \textit{a priori} knowledge of the curve solution. Namely we review and discuss aspects of transseries solutions \eqref{eq:transs-sol-nth} to physical inhomogeneous ODE of the type \eqref{eq:lin-nth-ODE} and, taken together with the section \ref{sec:Inner-outer} that follows, understand how to recover the same data (singularity structures $\Gamma_w(z)$, $\Gamma_z$ and germ expansions) as the curve solution.

As an example, let us consider the case of the second order ODE \eqref{eq:lin-2nd-ODE}. We seek a formal transseries solution
\begin{equation}
    y(z,\epsilon;\sigma_1,\sigma_2)=y^{(0)}(z,\epsilon)+\sigma_1 \mathrm{e}^{-\chi_1(z)/\epsilon}\,y^{(1)}(z,\epsilon)+\sigma_2\mathrm{e}^{-\chi_2(z)/\epsilon}\,y^{(2)}(z,\epsilon)\,.
\end{equation}
Each transseries sector $y^{(i)}(z,\epsilon)$ is an asymptotic series of the form:
\begin{equation}
    \label{eq:physical-yi-expansion}
    y^{(i)}(z,\epsilon)=\sum_{k\ge0} \epsilon^{k+1-\beta}\,y_k^{(i)}(z)\,.
\end{equation}
Substituting an ansatz $\mathrm{e}^{-\chi_i(z)/\epsilon}\,y^{(i)}(z,\epsilon)$ into the ODE, and assuming the asymptotic expansion above we obtain a set of recursion ODEs for the coefficients of the expansion. There is a completely determined particular solution with $\chi_0=0$ related to the inhomogeneous part \eqref{eq:lin-2nd-ODE}. The coefficients of its asymptotic expansion have $\beta=0$, and
\begin{eqnarray}
\label{eq:2ndODE-partsol-recursion}
    y^{(0)}_0 & = & \frac{R}{P_0}\quad;\quad y^{(0)}_1=-\frac{P_1}{P_0}\,(y_0^{(0)})'\\
    y^{(0)}_{k} & = & -\frac{1}{P_0}\left( P_1\,(y^{(0)}_{k-1})'+(y^{(0)}_{k-2})'' \right)\, ,\,\,k\ge 2\,.\nonumber
\end{eqnarray}
This is in full agreement with the germ ansatz in the Borel PDE, as expected. Using the map between transseries coefficients and Borel germ coefficients, $y_k^{(i)}\rightarrow\Phi_k^{(i)}\,\Gamma(k+1-\beta)$ we obtain the Borel germ recursion equation \eqref{eq:PDErecursion} and \eqref{eq:leading-coefficients}. The solution to these recursion equations is uniquely determined.

The homogeneous ODE will have further solutions with non-zero exponential weights $\chi_i(z)$. We obtain an algebraic equation for $\chi_i'(z)$ and recursion ODEs for the coefficients $y^{(i)}_k$. The respective value of $\beta$ is not fixed by the ODE, and the recursion ODEs for the $y^{(i)}_k(z)$ introduce a new integration constant for each $k$. These integration constants introduce an ambiguity in the determination of the coefficients of the asymptotic expansion. Furthermore, the algebraic equation also defines the $\chi_i$ up to another integration constant. Note that the exact same ambiguities appear from solving the Borel PDE. Considering again the case of the second order ODE \eqref{eq:lin-2nd-ODE} we obtain the following recursion ODEs for the exponentially suppressed transseries sectors ($i=1,2$):
\begin{eqnarray}
\label{eq:2nd-recursionODEs-gen}
    0 & = & (\chi_i'(z))^2-P_1(z)\,\chi_i'(z)+P_0(z) \nonumber\\
    0 & = & (y^{(i)}_0)'\,(2\chi_i'-P_1)+y^{(i)}_0\,\chi_i'' \\
    (y^{(i)}_{k-1})'' & = &  (y^{(i)}_k)'\,(2\chi_i'-P_1)+y^{(i)}_k\,\chi_i'' \, ,\,\,k\ge 1\,.\nonumber
\end{eqnarray}
Note that using the Borel map $y_k^{(i)}\rightarrow\Phi_k^{(i)}\,\Gamma(k+1-\beta)$ we obtain the equations for the coefficients of the Borel germs given by the Borel PDE. Solving the above equations clearly will introduce new integration constants per coefficient, and for the as of yet unknown position of the zeros of $\chi_i$. Nevertheless, the integration constants of the coefficients $y_k^{(i)}(z)$ can be fixed by a re-definition of the transseries coefficients $\sigma_i(\epsilon)$: they can be re-absorbed into an $\epsilon-$dependent series which multiplies the original series. We will see this shortly in an example. The unknown value of the constants in each $\chi_i$ will correspond to matching to the singularities of the coefficients of one of the other sectors (depending on which sector/node is connected to the one we are analysing---see e.g. figure \ref{fig:graphexample}). These singularities are elements of the set $\Gamma_z^{(j)}$ for $i \neq j$ defined section \ref{sec:algcurves}, and can be read from the recursion equation for the coefficients $y_k^{(0)},\,y_k^{(i)}$ in \eqref{eq:2ndODE-partsol-recursion} and \eqref{eq:2nd-recursionODEs-gen}. These ideas will be addressed more systematically in the following section.

\begin{example}
    Let us restrict our attention further to the differential equation with  $P_0=z$, $P_1=(1+z)$ and $R=-1$ first introduced in example \ref{eg:second}, solving the recursion equations for particular solution will return $y_k^{(0)}=\Gamma(k+1)\Phi_k^{(0)}$ with the germ coefficients $\Phi_k^{(0)}$ in \eqref{eq:coeffs-Phi0-curve} obtained from the curve expansion. Solving for the homogeneous sectors \eqref{eq:2nd-recursionODEs-gen}, we easily find $\chi_1'(z)=z$ and $\chi_2'(z)=1$, and
    \begin{eqnarray}\label{eq:physical-y1-coeffs}
            y_k^{(1)}(z) & = & \sum_{\ell=0}^k \frac{(-2)^{\ell-1/2} C^{(1)}_{k-\ell}}{(1-z)^{2\ell+1}}\,\Gamma(\ell+1/2)\,;\\
            y_k^{(2)}(z) & = & C_k^{(2)}\,,
            \label{eq:physical-y2-coeffs}
    \end{eqnarray}
    where the $C_k^{(i)}$ are unfixed integration constants introduced when solving the recursion ODEs. We may alternatively choose to package the integration constants as an unknown $\epsilon$-dependent asymptotic series that multiplies a fully determined asymptotic expansion as follows:
    \begin{eqnarray}
            y^{(1)}(z,\epsilon) & = & \sum_{k=0}^{+\infty}\sum_{\ell=0}^k \epsilon^k\,\frac{(-2)^{\ell-1/2} C^{(1)}_{k-\ell}}{(1-z)^{2\ell+1}}\,\Gamma(\ell+1/2)\nonumber\\
             & = & \left(\sum_{k=0}^{+\infty}C^{(1)}_{k}\epsilon^{k-1/2} \right)\, \sum_{\ell=0}^{+\infty}\frac{(-2)^{\ell-1/2} \epsilon^{\ell+1/2}}{(1-z)^{2\ell+1}}\,\Gamma(\ell +1/2)\, ; \\
            y^{(2)}(z,\epsilon) & = & \left(\sum_{k= 0}^{+\infty}C_k^{(2)}\epsilon^k\right)\,.
    \end{eqnarray}
    The coefficients on the second sum of $y^{(1)}(z,\epsilon)$ match the germ coefficients found from the curve \eqref{eq:coeffs-Phi1-curve} through a Borel map $\Phi_\ell^{(1)}\Gamma(\ell+1/2)$ (with the choice $\beta=1/2$). The coefficients of $y^{(2)}(z,\epsilon)$ are given by the overall unknown integration constants. This is expected if one looks at the singular germ at $\chi_1$: the singular behaviour of the coefficients is given by the difference $\chi_2-\chi_1$ (up to a choice of zeros), and its Borel germ can be summed to a simple pole at $w=\chi_2-\chi_1$, which corresponds to a truncated series consisting of a single constant term for $y^{(2)}(z,\epsilon)$.
\end{example} 

\begin{figure}
    \centering
    \includegraphics[scale=0.17]{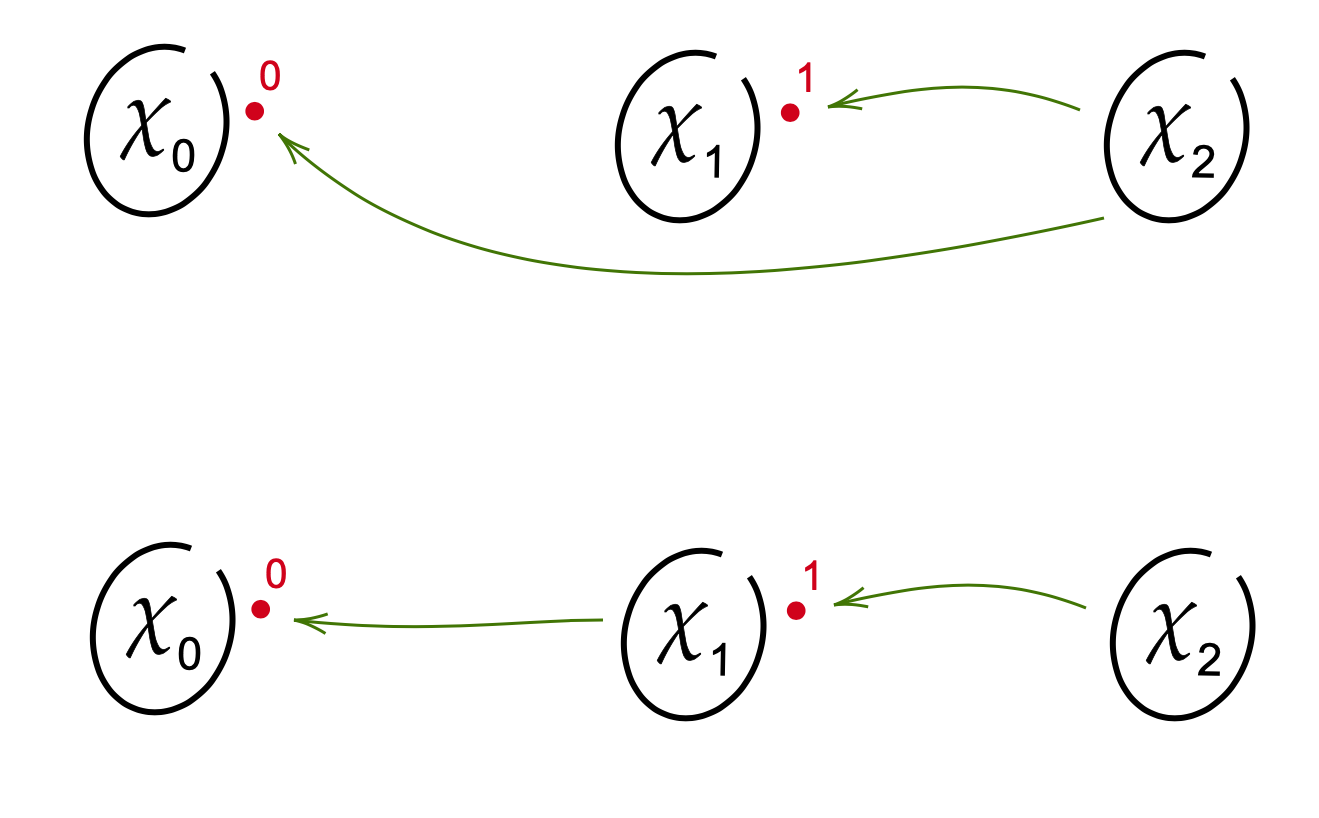}
    \caption{The two possible singularity graphs for the running example.}
    \label{fig:possiblesinggraphs}
\end{figure}

To summarise, in the language of section \ref{sec:algcurves}, from the above transseries ansatz we have so far learnt that 
\begin{equation}
    \hat{\Gamma}_w(z) = \{\chi_0, \chi_1, \chi_2\},
\end{equation}
with $\chi_0(z) = 0$ and $\chi_1'(z) = z$, $\chi_2'(z)=1$. We find that the corresponding $z$-singularity sets are
\begin{equation}
    \Gamma_z^{(0)} = \{z=0\},\quad \Gamma_z^{(1)} = \{z=1\}, \quad \Gamma_z^{(2)} = \emptyset \,.
\end{equation}
These can be determined immediately from \eqref{eq:2nd-recursionODEs-gen} since these equations are linear and we can read off the singular points from the coefficients. There are undetermined constants in $\chi_1$ and $\chi_2$ which give rise to the two possible singularity graphs $\Gamma$ illustrated in figure \ref{fig:possiblesinggraphs}. In the following section \ref{sec:Inner-outer} we discuss how this ambiguity, and the integration constants discussed above, may be fixed.

In the example above, we note that to agree with the analytic continuation given by the curve we simply set $C_0^{(1)}=1$ and all other integration constants to zero. However this need not be the case in general, and one might pose the question of how to identify the integration constants such that solution of the physical ODE matches the germ expansion obtained from analytic continuation in the Borel plane. We address this question systematically in the following section.

As previously remarked, a choice of integration constants corresponds to the multiplication of the physical solution $y^{(j)}(z,\epsilon)$ given by \eqref{eq:physical-yi-expansion} by an overall function $S^{(j)}(\epsilon)$ with expansion
\begin{equation}
\label{eq:stokes-exp}
    S^{(j)}(\epsilon)=\sum_{k\ge0}\,C^{(j)}_k \,\epsilon^k\,.  
\end{equation}
Then the re-summed product will be given by the convolution of corresponding Borel transforms:
\begin{equation}
   \mathcal{S}\left(S^{(j)}(\epsilon)\,y^{(j)}(z,\epsilon)\right) \simeq \int_0^{\infty}\mathrm{d}w\,\mathrm{e}^{-w/\epsilon}\,\left(C^{(j)}_0\,\Phi^{(j)}(w,z)+S_B^{(j)}(w)*\Phi^{(j)}(w,z)\right)\, ,
\end{equation}
where $\Phi^{(j)}(w,z)=\sum_{n\ge0}\Phi_n^{(j)}(z)\,w^{n-\alpha}$ is the Borel germ associated to the expansion $y^{(j)}(z,\epsilon)$, and we have introduced the Borel germ of $S^{(j)}(\epsilon)$ as 
\begin{equation}
S_B^{(j)}(w)=\sum_{n\ge0}\frac{C^{(j)}_{n+1}}{\Gamma(n+1)}\,w^n\,.
\end{equation}
Using the definition of the convolution and the germ expansion of $S_B^{(j)}(w)$ and $\Phi^{(j)}(w,z)$:
\begin{eqnarray}
    S_B^{(j)}(w)*\Phi^{(j)}(w,z) &\equiv& \int_0^{\infty}\mathrm{d}\tau\,S_B^{(j)}(w-\tau)\,\Phi^{(j)}(z,\tau) \\
        & = & \sum_{k\ge 1}\frac{w^{k-\alpha}}{\Gamma(k+1-\alpha)} \,\sum_{m=0}^{k-1}\,C^{(j)}_{k-m}
        \,\Gamma(m+1-\alpha)\,\Phi^{(i)}_m(z)\nonumber.
\end{eqnarray}

We would like to compare the above Borel plane convolution of our physical solution with unknown series $S^{(i)}(\epsilon)$ to the analytic continuation of the Borel germ from a previous singularity $\chi_i$ to $\chi_j$:
\begin{equation}
    \Phi^{(i\rightarrow j)}(w,z)=\sum_{n=0}^{+\infty}\Phi_n^{(i\rightarrow j)}(z)\,\left(w-(\chi_j(z)-\chi_i(z))\right)^{n-\alpha} 
\end{equation}
To do so we zoom into a point $z_{ij}$ with $\chi_j(z_{ij})-\chi_i(z_{ij})=0$. We obtain then
\begin{equation}
    \sum_{k\ge0}\,\Phi_k^{(i\rightarrow j)}(z_{ij})\,w^{k-\alpha}  =  C^{(j)}_0\,\Phi^{(j)}(z_{ij},w)+S_B^{(j)}(w)*\Phi^{(j)}(z_{ij},w)\,,
\end{equation}
which for equal powers of $w$ gives the coefficients of the multiplicative function $S^{(j)}(\epsilon)$
\begin{eqnarray}
    C_0^{(j)} & = & \frac{\Phi_0^{(i\rightarrow j)}(z_{ij})}{\Phi_0^{(j)}(z_{ij})}\,;\nonumber\\
   C^{(j)}_{k}  & = & \frac{\Gamma(k+1-\alpha)}{\Gamma(1-\alpha)}\,\frac{\Phi_k^{(i\rightarrow j)}(z_{ij})}{\Phi^{(i)}_0(z_{ij})}
    -\sum_{m=1}^{k}\,C^{(j)}_{k-m}
        \,\frac{\Gamma(m+1-\alpha)}{\Gamma(1-\alpha)}\,\frac{\Phi^{(i)}_m(z_{ij})}{\Phi^{(i)}_0(z_{ij})}\,.
\end{eqnarray}

\paragraph{Remark} We note that although the coefficients of the Borel germs are such that the series has a finite radius of convergence, the coefficients $C_m^{(j)}$ of the expansion $S^{(j)}(\epsilon)$ can be factorially divergent, making the expansion \eqref{eq:stokes-exp} asymptotic.

\section{Inner-outer matching}
\label{sec:Inner-outer}

In this section we focus on the resolution of the ambiguities that appear in the transseries analysis. We resolve these systematically through an inner-outer germ matching structure: each transseries sector is defined as the analytic continuation of a Borel germ associated to a sector connected by an arrow in the singularity graph in figure \ref{fig:graphexample}. We begin with the general complex analytic structure and then turn to computing the transseries in our running example \ref{eg:runninggerms}.

\subsection{Germ matching}
Let us consider the singularity graph in figure \ref{fig:graphexample}. We consider a subset of two nodes joined by a single arrow. Associated to this data are two germs which coalesce at a point $z=z_{\star}$. In this section we discuss how to \textit{match} one germ onto another. That is, the germ $\Phi^{(1)}(w,z)$ is specified and we seek to determine the analytic continuation to germ $\Phi^{(2)}(w,z)$.

We label the two nodes of the graph by $\chi_1$ and $\chi_2$ and write the associated (singular) germs as
\begin{equation}
    \label{eq:sing-germs-12}
    \Phi^{(i)}(w,z) = (w-\chi_i(z))^{-\alpha_i}\,\,\sum_{k\ge0}\,\Phi^{(i)}_k(z)\,(w-\chi_i(z))^k
 + \text{reg.}\,,\quad i=1,2\,.
\end{equation}
These germs correspond to the transseries sectors $y^{(i)}(z,\epsilon)$ in \eqref{eq:transs-sol-nth}, with coefficients given by
\begin{equation}
\label{eq:y-i-expn}
    y^{(i)}(z,\epsilon)\simeq \epsilon^{1-\alpha_i}\,\sum_{n\ge0}\,\epsilon^n\,y_n^{(i)}(z)\,,\quad y_n^{(i)}(z)=\Phi_n^{(i)}\,\Gamma(n+1-\alpha_i)\,,\quad i=1,2\,.
\end{equation}

The inner-outer matching procedure works under a number of assumptions that we outline below. We will see in the following that these assumptions are satisfied for our running example and we expect them to be satisfied for a wide class of differential equations.

We suppose that $\chi_{12}(z):=\chi_2(z)-\chi_1(z)$ has a zero of order $\beta_{12}$ at $z=z_{12}$ and write
\begin{equation}
    \chi_{12}(z) \sim (z-z_{12})^{\beta_{12}} + \text{h.o.t.}\,,
\end{equation}
where, for simplicity, we suppose $\beta_{12}$ is a positive integer. We further suppose that the coefficients have singularities in $z$ scaling as
\begin{equation}\label{eq:leading-z-singular-behaviour}
    \Phi_n^{(1)} \sim (z-z_{12})^{-\gamma_1 - n\Delta_1} \,,\quad \Phi_n^{(2)} \sim (z-z_{12})^{-\gamma_2 - n \Delta_2}\, ,
\end{equation}
for some rational numbers $\gamma_1,\,\gamma_2$ and $0\le\Delta_1,\Delta_2\le\beta_{12}$. This singularity growth is expected generally in singularly perturbed ODEs assuming a singularity of the first coefficient $\Phi^{(i)}_0\sim (z-z_{12})^{-\gamma_i}$. Without loss of generality we assume $\Delta_1=\Delta_2=\beta_{12}$, as the case where the singular behaviour of the coefficients grows at a lower rate (\textit{i.e.}  $0<\Delta_i<\beta_{12}$) is included in the analysis below, with the first expansion coefficients zero. Furthermore, note that if this analysis was done between two nodes in the singularity graph that are not connected, the growth of the coefficients would not agree with the order of the relevant zero.

We will see that for a consistent inner-outer germ matching procedure we further require
\begin{equation}\label{eq:consistency-inner-outer}
    -\gamma_1 -\alpha_1\beta_{12} = -\gamma_2 -\alpha_2\beta_{12}\,.
\end{equation}

\paragraph{Inner variables.}
Associated to the relevant arrow in the matching graph of figure \ref{fig:graphexample}, we associate an inner variable
\begin{equation}
    s_{12} = \frac{w-\chi_1(z)}{\chi_2(z)-\chi_1(z)}\,.
\end{equation}
To ease notation, and where no confusion arises, we write $s=s_{12}$ in the following. We also note that performing this change of variables in the Borel planes corresponds to changing the parameter $\epsilon$ in the physical plane to $\delta=\epsilon/\chi_{12}(z)$.

Let us now re-express both germ expansions in terms of the inner variable:
\begin{equation}
    \Phi^{(1)}(s,z) = s^{-\alpha_1}\,\,\sum_{k\ge0}\,\varphi_k^{(1)}(z)\,s^k+\text{reg.},
    \label{eq:inner-expansion-zero}
\end{equation}
and
\begin{equation}
    \Phi^{(2)}(s,z) = (s-1)^{-\alpha_2}\,\,\sum_{k\ge0}\,\varphi_k^{(2)}(z)\,(s-1)^k +\text{reg.}\,,
    \label{eq:inner-expansion-one}
\end{equation}
where we have also introduced new $z$-dependent re-scaled coefficients defined as
\begin{equation}
\label{eq:coeffs-varphi}
    \varphi_n^{(i)}(z) := (\chi_2(z)-\chi_1(z))^{-\alpha_{i} + n} \,\,\Phi_n^{(i)}(z) ,\,\,i=1,2\,.
\end{equation}
The assumptions discussed above then give rise to the following scaling behaviour of these coefficients as $z \to z_{12}$:
\begin{equation}
    \label{eq:sing-varphi}
    \varphi_n^{(i)}(z) \sim (z-z_{12})^{-\gamma_i - \alpha_i \beta_{12}}\,,\quad i=1,2.
\end{equation}
We note that all of the above scaling relations are a `not more singular than' and examples can (and indeed will) not always saturate these scalings.

The inner-outer germ matching procedure reorders the expansions in $z$ and $s$ making crucial use of Darboux's theorem \cite{dingle1973asymptotic} to relate the late terms of an expansion at a regular point $s_0$ with the coefficient of the singular expansion around its nearest singularity $s_1$. Taking the above expansions close to $z=z_{12}$ ensures that $\chi_1$ and $\chi_2$ are the nearest singularities, and the transformation to the associated inner variable ensures that these singularities are separated at a fixed distance in the $s-$plane. More precisely, we are interested in relating the following regular expansion at $s=0$:
\begin{equation}
    \Phi^{(1)}(s,z)\, s^{\alpha_1} = \sum_{k\ge0}\,\varphi_k^{(1)}(z)\,s^k\,,
\end{equation}
and the singular expansion at $s=1$:
\begin{equation}
    \Phi^{(2)}(s,z)\,s^{\alpha_1}  =  (s-1)^{-\alpha_2}\,\,\sum_{k\ge0}\,(s-1)^k\,\sum_{n=0}^k\,\,\frac{\Gamma(\alpha_1+1)}{\Gamma(\alpha_1+1-n)\,n!}\,\,\varphi_{k-n}^{(2)}(z)
    \,,
\end{equation}
 where here we have expanded the factor $s^{\alpha_1}$ at $s=1$. Darboux's theorem gives an asymptotic relationship between the coefficients: 
 \begin{equation}
     \varphi_{k\gg1}^{(1)}\sim \frac{\Gamma(k+\alpha_2)}{(-1)^{\alpha_2}\Gamma(\alpha_2)\,k!}\,\sum_{r\ge0}\,\frac{\Gamma(\alpha_2)\Gamma(k+\alpha_2-r)}{\Gamma(k+\alpha_2)\Gamma(\alpha_2-r)}(-1)^r\,\,\sum_{n=0}^{r}\,\frac{\Gamma(\alpha_1+1)}{\Gamma(\alpha_1+1-n)\,n!}\,\varphi_{r-n}^{(2)}(z)\,.
 \end{equation}

\paragraph{Inner expansion.}
The inner variable allows us to switch the order of expansion. Rather than expanding about $\chi \in \Gamma_w$ points we expand about the relevant point $z=z_{\star}$ in $\Gamma_z^{(\chi)}$.

We therefore consider the following `inner' expansions of the $\varphi_n^{(i)},\,i=1,2$ coefficients organised into the following grid ($i=1,2$)
\begin{equation}
\begin{split}
    \varphi_0^{(i)}(z) &= (z-z_{12})^{-\gamma_i - \alpha_i \beta_{12}}\,\left( \psi_{0,0}^{(i)} + (z-z_{12}) \,\psi_{0,1}^{(i)} + (z-z_{12})^2\, \psi_{0,2}^{(i)} +\ldots \right) \,;\\
    \varphi_1^{(i)}(z) &= (z-z_{12})^{-\gamma_i - \alpha_i \beta_{12}}\,\left( \psi_{1,0}^{(i)} + (z-z_{12}) \,\psi_{1,1}^{(i)}+ (z-z_{12})^2 \,\psi_{1,2}^{(i)} + \ldots \right) \,;\\
    \vdots
    \\
\end{split}    
\end{equation}
Comparing order-by-order, we obtain an infinite series (for each power of $(z-z_{12})$) of \textit{constant} resurgence/Borel problems, where the singularity location is fixed to $s=1$. We may then apply late-orders analysis for each $m=0,1,2,\ldots$. Darboux's theorem yields the following large $n$ relation
\begin{eqnarray}
    \hspace{-20pt} (-1)^{\alpha_2}\,\psi_{k,m}^{(1)} & \sim & \frac{\Gamma(k+\alpha_2)}{\Gamma(\alpha_2)\,k!}\,\sum_{r\ge0}\,\frac{\Gamma(\alpha_2)\Gamma(k+\alpha_2-r)}{\Gamma(k+\alpha_2)\Gamma(\alpha_2-r)}(-1)^r\,\,\sum_{n=0}^{r}\,\frac{\Gamma(\alpha_1+1)}{\Gamma(\alpha_1+1-n)\,n!}\,\psi_{r-n,m}^{(2)}\nonumber \\
     & \sim & \frac{\Gamma(k+\alpha_2)}{\Gamma(\alpha_2)\,k!}\,\left(\psi_{0,m}^{(2)} - \frac{\alpha_2-1}{k+\alpha_2 -1}\,(\psi_{1,m}^{(2)}+\alpha_1\,\psi_{0,m}^{(2)}) + \ldots\right)\,.
     \label{eq:Darboux-for-psis}
 \end{eqnarray}

The key idea is as follows. From the large order asymptotics of the \textit{known} coefficients $\psi_{k,m}^{(1)}$ we may determine the early orders of the unknown $\psi_{0,m}^{(2)},\psi_{1,m}^{(2)},\psi_{2,m}^{(2)},\ldots$ at each order in $m=0,1,2\ldots\,$. In the case where $\Psi$ arises as the solution to a differential equation, these constants will yield initial data for the \textit{outer} differential equation problems.

This procedure is called \textit{inner-outer matching}. In general, the task is then to repackage these new coefficients and re-expand/analytically continue\footnote{The analytic continuation can equivalently be obtained from the underlying ODE, if there is one, using methods such as Padé approximants, or Darboux's theorem to link the expansion of the function $\varphi^{(2)}_k(z)$ at neighbouring singularities $z_{12}$ and $z_{23}$.} to a new singular point   $z_{23}$ of the coefficients $\Phi_n^{(2)}$ in \eqref{eq:sing-germs-12}. This new singularity would now connect singular germs $\Phi^{(2)}(w,z)$ and $\Phi^{(3)}(w,z)$ and in this way one may iterate through a full singularity graph. We demonstrate the procedure in our running example below.

\subsection{Stokes constants}

Let us focus on how the collections of coefficients $\{\psi\}$ may be understood as coefficients of some \textit{parameter}-dependent Stokes constants of the inner problem. We revisit the transseries sectors associated to the subset in figure \ref{fig:graphexample} connecting transseries sectors $y^{(1)}(z,\epsilon)$ and $y^{(2)}(z,\epsilon)$ in \eqref{eq:transs-sol-nth} with expansions of the form \eqref{eq:y-i-expn}. We first re-write this subset of the transseries as 
\begin{equation}
    \mathrm{e}^{-\chi_1(z)/\epsilon}\,\sigma_1\left( y^{(1)}(z,\epsilon)+\frac{\sigma_2}{\sigma_1}\,\mathrm{e}^{-\chi_{12}(z)/\epsilon}\,y^{(2)}(z,\epsilon)
    \right)\,.
\end{equation}

Recall that the inner-outer matching analysis of the previous subsection allowed us to determine the coefficients of the sector $y^{(2)}(z,\epsilon)$ from a known sector $y^{(1)}(z,\epsilon)$. This is related to the introduction of a new physical parameter, associated to the inner-Borel variable $s$, given by
 \begin{equation}
 \delta = \epsilon / \chi_{12}\,,
 \end{equation}
 whereby the transseries subset may be organised as\footnote{We may discard the overall factors, as they play no role in the Stokes switching of $y^{(2)}$ by the resummed $y^{(1)}$ crossing a $\ell_{12}$ Stokes line.}
 \begin{equation}
 \label{eq:inner-y1-y2}
     y^{(1)}(z,\delta)+\bar{\sigma}\,\mathrm{e}^{-1/\delta}\,y^{(2)}(z,\delta)\,,
\end{equation}
with $\bar{\sigma}=\sigma_2/\sigma_1$ and the expansion of the transseries sectors $y^{(i)}$ in the new parameter $\delta$ is simply
\begin{equation}
    \chi_{12}(z)\,y^{(i)}(z,\delta)\sim\sum_{k\ge0}\delta^{k+1-\alpha_i}\,\varphi_k^{(i)}(z)\,\Gamma(k+1-\alpha_i)\,,\quad i=1,2\,.
\end{equation}
The coefficients $\varphi_k^{(i)}(z)$ are those introduced in the previous subsection \eqref{eq:coeffs-varphi}. 

We may then analyse the Stokes phenomena associated to this inner problem in the parameter $\delta$. Through the inner-outer matching procedure presented above we can determine the analytic continuation of the Borel germ of $y^{(1)}(z,\delta)$ at its Borel singularity $s=1$ (or in original variable at $w=\chi_{12}(z)$), which unambiguously determines the coefficients of $y^{(2)}(z,\delta)$. On the other hand, we have seen that each transseries sector can be determined from the defining ODE up to an $\epsilon$-dependent factor, and this translates\footnote{In the previous section it was mentioned that the singular behaviour of the $\varphi_k^{(i)}(z)$ when $z\rightarrow z_{12}$ cannot be worse than \eqref{eq:sing-varphi}. In the case some of these coefficients have the leading coefficients zero  $\psi_{n,m<m_n}=0$, then fixing the ambiguity in the transseries sector corresponds to choosing the value of first non-zero coefficient $\psi_{n,m_n}$ for each $n$.} to different choices of the coefficients $\psi_{n,0}^{(2)}$. Assume a choice of normalisation of the sector $y^{(2)}$ such that
\begin{equation}
   \left. (z-z_{12})^{\gamma_2+\alpha_2\,\beta_{12}}\,\chi_{12}(z)\,\tilde{y}^{(2)}(z,\delta)\right|_{z=z_{12}} = 1+\mathcal{O}(z-z_{12})\,.
\end{equation}
Expanding $\tilde{y}^{(2)}(z,\delta)$ in $\delta$ and $z-z_{12}$ gives rise to expansion coefficients $\tilde{\psi}_{n,m}^{(2)}$ normalised as $\tilde{\psi_{n,0}}=\delta_{n,0}$. We may then write the relation between the analytically continued $y^{(2)}(z,\delta)$, as obtained through the inner-outer matching, and the normalised expansion
as
\begin{equation}
y^{(2)}(z,\delta)=S(\delta)\;\tilde{y}^{(2)}(z,\delta)\,,
\label{eq:relation-norm-ac}
\end{equation}
where
\begin{equation}
    S(\delta)=\sum_{k\ge0}\,\delta^k\,\psi_{k,0}^{(2)}\,.
\end{equation}
This is straightforward to verify by multiplying \eqref{eq:relation-norm-ac} by $(z-z_{12})^{\gamma_2+\alpha_2\,\beta_{12}}$ and evaluating at $z=z_{12}$.

To summarise, we may solve the effective ODEs for $y^{(2)}$ in the inner variables $z,\delta$ with the simple initial data above and the $\{\psi\}$ gain an interpretation as Stokes constants which may obtained from the matching procedure outlined above. One thus has two options: either obtain the $\{\psi\}$'s from the matching procedure, or solve the ODEs with trivial initial data and perform a convolution/multiplication as above.

However this data is obtained, the Stokes automorphism describing the switching of sector $y^{(2)}$ by crossing the Stokes lines emanating from sector $y^{(1)}$ is then given by a \textit{constant} jump of the parameter $\bar{\sigma}$ in \eqref{eq:inner-y1-y2} (independent of $\delta$). The same conclusion then holds for the original $\epsilon$ transseries variable.

\subsection{Example}
We can carry out the above procedure for our running example, starting from the Borel PDE \eqref{eq:running-example-PDE}. The germ solution at $w=0$ is given by \eqref{eq:sing-germs-12} with $i=0$, $\alpha_0=0$, $\chi_0=0$:
\begin{equation}
    \Phi^{(0)}(w,z)=\sum_{k\ge0}\,\Phi_k^{(0)}(z)\,w^k\,,
\end{equation}
and the coefficients $\Phi_k^{(0)}$ can be determined via recursion relations, obeying the initial conditions provided by the inhomogeneous term in the original ODE \eqref{eq:running-example-ODE}, and are given in \eqref{eq:coeffs-Phi0-curve}. We would like to construct the analytic continuation of this germ around the two Borel singularities $\chi_1=z^2/2+p_1$ and $\chi_2=z+p_2$. These singularities can be determined up to a constant ($p_i$) from the Borel PDE, and this constant can be determined from the singular behaviour of the coefficients $\Phi_k^{(0)}$ of the germ at $z=0$:
\begin{equation}
    \Phi_k^{(0)}\sim z^{-1-2k}=z^{-\gamma_0-\Delta_0 k}\,,
\end{equation}
where $\gamma_0=1,\,\Delta_0=2$. 

\subsubsection{Node selection}

There are now two possible direct links between nodes :
\begin{itemize}
\item $0\leftarrow 1$: The Borel singularity is given by $\chi_{01}=\chi_1-\chi_0=z^2/2$ where we have taken $p_1=0$ such that $\chi_{01}$ has a zero at $z=0$, the singular point of the coefficients $\Phi_k^{(0)}$. The order of the zero is $\beta_{01}=2$, which needs to obey $\Delta_0\le\beta_{01}$. These two nodes are then connected.

\item $0\leftarrow 2$: The Borel singularity is given by $\chi_{02}=\chi_2-\chi_0=z$ where we now would take $p_2=0$ such that $\chi_{02}$ has a zero at $z=0$. The order of the zero is $\beta_{02}=1$, which does not obey $\Delta_0\le\beta_{02}$. Thus there is no direct link $0\leftarrow 2$.
\end{itemize}

This analysis shows that the links we will have are given on the bottom singularity graph of figure \ref{fig:possiblesinggraphs}. Thus we need to perform the inner analysis of the link $0\leftarrow 1$.

\subsubsection{Singular behaviour}

The germ expansion around $w=\chi_{1}(z)$ can also be obtained from the Borel PDE and has the form given in \eqref{eq:sing-germs-12} with $i=1$, where the value of $\alpha_1$ gets determined by the singular behaviour of the coefficients $\Phi_k^{(1)}(z)$ at $z=0$. From \eqref{eq:leading-z-singular-behaviour} We expect the coefficients to behave as 
\begin{equation}
    \Phi_k^{(1)}\sim z^{-\gamma_1-k\Delta_1}\,,
\end{equation}
where for a consistent inner-outer matching we need $1=\gamma_0+\alpha_0\beta_{01}=\gamma_1+ \alpha_1\beta_{01}=\gamma_1+ 2\alpha_1$. The values of $\gamma_1,\Delta_1$ can be obtained from the singular behaviour of the first coefficients $\Phi_k^{(1)}(z), k=0,1$ at $z=0$. This can be found directly from the Borel PDE, and gives $\gamma_1=0=\Delta_1$ (\textit{i.e.} the coefficients of this germ are non-singular at $z=0$). We find $\alpha_1=1/2.$
%

\subsubsection{Inner analysis}

We now have all the ingredients to perform the inner analysis. The inner variable in this case is
\begin{equation}
    s_{01}=\frac{w-\chi_0}{\chi_{1}-\chi_0}=\frac{w}{\chi_1(z)}.
\end{equation}
In terms of this variable we can re-write our Borel germs as in \eqref{eq:inner-expansion-zero} (with $\alpha_0=0$) and \eqref{eq:inner-expansion-one}, where the coefficients given in \eqref{eq:coeffs-varphi} admit an expansion around the singular point $z=0$ of the form:
\begin{equation}\label{eq:running-ex-exp-zero}
    \varphi_k^{(i)}(z)=z^{-1}\,\sum_{m\ge0}\psi_{k,m}^{(i)}\,z^m\quad,\,i=0,1.
\end{equation}
The coefficients $\psi_{k,m}^{(0)}$ can be unambiguously determined via recursion relations from \textit{e.g.} the Borel PDE and inhomogeneous condition defining the first two terms, and in this simple case we obtain
\begin{eqnarray}
    \psi_{k,2m}^{(0)} & = & -\frac{1}{\sqrt{\pi}}\frac{\Gamma(k-m+1)\,\Gamma(k-m+1/2)}{\Gamma(k-2m+1)\,\Gamma(k+1)}\,,m=0,\cdots,\left\lfloor\frac{k}{2}\right\rfloor;\nonumber\\
     \psi_{k,2m+1}^{(0)} & = & -\frac{1}{\sqrt{\pi}}\frac{\Gamma(k-m)\,\Gamma(k-m+1/2)}{\Gamma(k-2m)\,\Gamma(k+1)}\,,m=0,\cdots,\left\lfloor\frac{k-1}{2}\right\rfloor.
     \label{eq:running-ex-psi-coeffs-inner}
\end{eqnarray}
The coefficients $\psi_{k,m}^{(1)}$ will be determined up to integration constants, as was discussed in section \ref{sec:Outer}. Note that while in this example we have the exact coefficients for all $k$, in a more general case we would be able to determine the large $k\gg1$ behaviour of the coefficients by determining a large number of them recursively and using extrapolation methods such as Richardson transforms \cite{Aniceto:2018bis}.

To write the relationship between coefficients at $s=0$ and at its nearest singularity $s=1$, we first re-write the Borel PDE \eqref{eq:running-example-PDE} in the inner variables $z,s$. We obtain $\mathscr{P}_B\,\Phi^{(0)}(s,z)$ with operator
\begin{eqnarray}              
\mathscr{P}_B & = & z^3\partial_z^2+\left(2z(1+z)-4z^2s\right)\,\partial_z\partial_s- \\
&  & -\left(4(1+z)-6z\,s\right)\partial_s+4\left(1-(1+z)s+z\,s^2\right)\partial_s^2\,.\nonumber
\end{eqnarray}
The Borel PDE then translates to the recursion relations for $\ell\ge-1$
\begin{eqnarray}
   0& = & ((\ell-1)(\ell-2)+2k(2k-2\ell+3))\psi_{k,\ell}^{(0)}+2(k+1)((\ell-1)-2(k+1))\psi_{k+1,\ell}^{(0)}+\nonumber\\
    & &+2(k+1)(\ell-2(k+1))\psi_{k+1,\ell+1}^{(0)}+4(k+1)(k+2)\psi_{k+2,\ell+1}^{(0)},
\end{eqnarray}
where $\psi_{k,-1}^{(0)}=0$. We now use the the relation between the expansion of the Borel transform at $s=0$ and $s=1$ from Darboux's theorem, which related large order coefficients  $\psi_{k,m}^{(0)}$ with $\psi_{k',m}^{(1)}$ for small $k'$, as given in \eqref{eq:Darboux-for-psis}. For our example we find that for $k\gg1$
\begin{equation}
    \,\psi_{k,m}^{(0)}\sim-\frac{S_{01}}{2}\frac{(-1)^{-\frac{1}{2}}\Gamma(k+1/2)}{\Gamma(1/2)\Gamma(k+1)}\,\left(\psi_{0,m}^{(1)}+\frac{\psi_{1,m}^{(1)}}{2k}+\frac{3\psi_{2,m}^{(1)}+\psi_{1,m}^{(1)}}{4k^2}+\cdots\right)\,,
\end{equation}
with the Stokes constant $S_{01}=2$ determined in \eqref{eq:piecewise-S01}. 
Thus we can determine the coefficients $\psi_{k,m}^{(1)}$, and construct the analytically continued germ at $s=1$, or equivalently at $w=\chi_1$, as a Taylor series in $z$. For our particular example, the knowledge of the coefficients for the germ at $s=0$ in closed form \eqref{eq:running-ex-psi-coeffs-inner}, allows us to determine any coefficient of the analytic continuation at $s=1$. From \eqref{eq:running-ex-psi-coeffs-inner} and Darboux's theorem we can easily see that $\psi_{0,0}^{(1)}=\psi_{0,1}^{(1)}=(-1)^{\frac{1}{2}}$ and
\begin{equation}
    (-1)^{\frac{1}{2}}\prod_{\ell=1}^{m}\frac{k+1-\alpha-m-\ell}{k+\frac{1}{2}-\ell}  
    = \psi_{0,2m+\alpha}^{(1)}+\sum_{r\ge1}\,(-1)^r\psi_{r,2m+\alpha}^{(1)}\,\prod_{\ell=1}^{r}\frac{\frac{1}{2}-\ell}{k+\frac{1}{2}-\ell}\,, 
\end{equation}
for $m\ge1$, $\alpha=0,1$. We can write the l.h.s. of the equations above as 
\begin{equation}
    \prod_{\ell=1}^{m}\frac{k+1-\alpha-m-\ell}{k+\frac{1}{2}-\ell}=1+\sum_{r=1}^{m}\frac{a^{(\alpha)}_{m,r}}{\prod_{\ell=1}^{r}\left(k+\frac{1}{2}-\ell\right)}\,,\,\alpha=0,1,
\end{equation}
where 
\begin{equation}
    \prod_{\ell=1}^{m}(k+1-\alpha-m-\ell)=a^{(\alpha)}_{m,m}+\sum_{\ell=0}^{m-1}a^{(\alpha)}_{m,\ell}\,\prod_{s=\ell}^{m-1}\left(k-\frac{1}{2}-s\right)\,.
\end{equation}
It is now evident that we can make the following identification:
\begin{eqnarray}
    \psi_{0,m}^{(1)}&=&(-1)^{\frac{1}{2}};\\
    \psi_{r,2m+\alpha}^{(1)} &=& \frac{(-1)^{r+\frac{1}{2}}\,a_{m,r}^{(\alpha)}}{\prod_{\ell=1}^{r}(\frac{1}{2}-\ell)}\,,\hspace{15pt}\alpha=0,1\,,r\le m\,;\\
    \psi_{r,2m+\alpha}^{(1)} &=& 0\,,\hspace{70pt}\alpha=0,1\,,\,r>m\,.
\end{eqnarray}
This is enough information to fix all unknown constants that appear in the direct calculation of the germ at $w=\chi_1$, given by the solution of the ODE (or equivalently the Borel PDE) \eqref{eq:physical-yi-expansion}, with coefficients $y_k^{(1)}$ given by \eqref{eq:physical-y1-coeffs}. The Borel germ at $w=\chi_1$ (with a choice of $\beta=1/2$ to cancel the factorial growth) is then
\begin{equation}
    \Phi^{(1)}(w,z)=\sum_{k\ge 0}(w-\chi_1)^{k-\frac{1}{2}}\frac{y_k^{(1)}}{\Gamma(k+\frac{1}{2})}.
\end{equation}
In inner variables $\Phi^{(1)}(z,s)$ will have an expansion of the form \eqref{eq:inner-expansion-one}, with $\alpha_2=1/2$ and coefficients 
\begin{equation}
\varphi_k^{(1)}(z)=\frac{\chi_1^{k-\frac{1}{2}}y_k^{(1)}}{\Gamma(k+\frac{1}{2})}\,.
\end{equation}
These coefficients can be expanded at $z=0$ as in \eqref{eq:running-ex-exp-zero}. We can then use the coefficients $\psi_{k,m}^{(1)}$ calculated above via the inner analysis to determine the integration constants $C_\ell^{(1)}$ of the $y_k^{(1)}$ in \eqref{eq:physical-y1-coeffs}: we obtain $C_0^{(1)}=1$ and $C_{\ell\ge1}^{(1)}=0$, as expected by the analytic continuation given by the algebraic curve description.

\subsubsection{Connecting nodes \texorpdfstring{$1\leftarrow 2$}{1to2}}
Once the integration constants of the Borel germ at $w=\chi_1$ are known we can perform the previous analysis to connect the Borel germs $\Phi_1(w,z)$ and $\Phi_2(w,z)$. The coefficients 
\begin{equation}
\Phi_k^{(1)}=\frac{(-2)^{k-1/2}}{(1-z)^{2k+1}}\sim(z-z_{12})^{-\gamma_1-k\Delta_1}
\end{equation}
have a singular behaviour at $z=z_{12}=1$, with $\gamma_1=1$ and $\delta_1=2$. The germ $\Phi_2(w,z)$ will also have an expansion like \eqref{eq:sing-germs-12}, where the coefficients have an expected singular behaviour at $z=z_12$ of the type \eqref{eq:leading-z-singular-behaviour}. However from the ODE we know that all $\Phi_k^{(2)}=C_k^{(2)}$ are constant (see \eqref{eq:physical-y2-coeffs}), and thus $\gamma_2=\Delta_2=0$.

The singular behaviour of the coefficients $\Phi_k^{(1)}$ is related to the zero of $\chi_{12}=\chi_2-\chi_1$. The undetermined constant of the Borel singularity $\chi_2$ is the determined such that $\chi_{12}=0$ at $z=z_{12}=1$, and we find $\chi_2=z-1/2$ and
\begin{equation}
    \chi_{12}(z)\equiv \chi_2-\chi_1=-\frac{(z-1)^2}{2}.
\end{equation}
This corresponds to a second order zero $\beta_{12}=2$. Using the consistency condition \eqref{eq:consistency-inner-outer} one finds $\alpha_2=1$ and we expect 
\begin{equation}
    \Phi_2(w,z)=\frac{\Phi_0^{(2)}}{w-\chi_2}+\mathrm{reg.}\,.
\end{equation}
To determine the analytic continuation of the Borel germ at $w=\chi_2$, $\Phi_1(w,z)$, to $w=\chi_2$, we use the inner variable $s=(w-\chi_1(z))/\chi_{12}(z)$, and now expand the coefficients of expansions \eqref{eq:inner-expansion-zero} and \eqref{eq:inner-expansion-one} in powers of $(z-1)$ 
\begin{eqnarray}
\varphi_k^{(1)}(z) & \equiv & \Phi_k^{(1)}\chi_{12}^{k-1/2}=-(1-z)^{-2}\,;\\
\varphi_k^{(2)}(z) & \equiv & \Phi_k^{(2)}\chi_{12}^{k-1}=(1-z)^{2k-2}\,\frac{C_k^{(2)}}{(-2)^{k-1}}\,.
\end{eqnarray}
It is now straightforward to use Darboux's theorem \eqref{eq:Darboux-for-psis}\footnote{We again need to take into account the overall factor $-\frac{S_{12}}{2\pi\mathrm{i}}$, where $S_{12}=-\pi\mathrm{i}$ is given in \eqref{eq:piecewise-S12}.}, and directly obtain $C_\ell^{(2)}=\delta_{\ell,0}(-1)^{1/2}$.

This result is expected as we are analysing the analytic continuation of the singular part of the Borel germ at $w=\chi_1$ at the singularity $w=\chi_2$. This exactly corresponds to expanding the singular part of the germ \eqref{eq:curve-Phi1-germ} at $w=\chi_2$.

\begin{remark}
We conclude with some remarks on how the above inner outer matching procedure relates to the curve solution $P_z=0$. Let us consider matching germ $\chi_0$ to $\chi_1$ via $z=0$ with the associated inner variable $s=w/\chi_1(z)$. The re-scaled curve becomes
\begin{equation}    
    P_z(s,\varphi) = (s-1)(sz^4 - 2z^3 +z^2)\varphi_B^2 + 2(s-1)z^2 \varphi_B + 1 = 0
\end{equation}
The inner-outer matching is associated to an arrow in the graph of figure \ref{fig:graphexample} so we also have a distinguished point $z=0 \in \Gamma^{\chi_1}_z$. The Borel solution is then expanded near $z=0$ as
\begin{equation}
    \varphi_B(s,z) = \frac{1}{z} \varphi_0(s) + \varphi_1(s) + \varphi_2(s)z + \ldots
\end{equation}
Substituting this expansion into $P_{z}(s,\varphi)=0$ and collecting powers of $z$ we obtain a family of algebraic curves for the $\varphi_i(s)$:
\begin{equation}
\begin{split}
    (s-1)\varphi_0(s)^2 + 1 &= 0 \\
    \varphi_0(s) - \varphi_1(s) - 1  &= 0 \\
    \ldots
\end{split}    
\end{equation}
The inner-outer matching is consistent since each of these has singularities only at $s=0$ and $s=1$ and the pole singularity is at infinity.
\end{remark}

\section{Summary and further directions}

In this work, we have developed a novel method for generating pairs of algebraic curve solutions and associated ordinary differential equations, providing exact solutions within this framework. By associating transseries with these algebraic curves, we were able to investigate the conformal (exponential) relationship between $\epsilon$ and $1/n$ expansions and gain insight into the higher-order Stokes phenomenon. Additionally, we developed a complex analytic inner-outer matching structure that simplifies the treatment of parametric resurgence. This approach reduces parametric resurgence to a `power series' of constant resurgence problems whereby the introduction of an inner variable allows one to zoom in on the coalescence of two singularities, keeping their distance fixed in the inner $s$-plane.

\paragraph{Further directions.}
Through the examples presented here, we have made a first step to formulating notions of co-equational resurgence in a geometric setting. On the route to full parametric resurgence many interesting questions remain and we conclude with a brief discussion of a selection of these.

We would like to extend the present work to examples of algebraic curves $\Sigma_{z_1,z_2}$ with two moduli $z_1$ and $z_2$ that arise as solutions to singularly perturbed \textit{partial} differential equations. In this case the singularity sets $\Gamma_{z_1,z_2}$ now have non-trivial geometry and should be treated more carefully as divisors on $\mathbb{C}_{z_1}\times \mathbb{C}_{z_2}$ rather than isolated points. The inner-outer matching procedure will correspondingly be considerably richer.

In section \ref{subsec:automorphisms} we discussed an interesting relationship between the Stokes automorphism group associated to a transseries and the Stokes automorphism group associated to the large $n$ transseries expansion of the transseries components themselves. This structure gives rise to Stokes phenomena of Stokes constants and piecewise constant behaviour. It would also be interesting to investigate this relationship further, in particular to develop the associated $1/n$ bridge equations and Alien calculus. A further avenue of study is the connection between the $1/n$ Stokes automorphism and the Stokes smoothing recently developed in \cite{howls2024smoothing}.

\paragraph{Acknowledgements.}
The authors conducted part of this research while visiting the Okinawa Institute of Science and
Technology (OIST) and are grateful for the support of the Theoretical Sciences Visiting Program (TSVP) and Reiko Toriumi. SC would also like to thank Masazumi Honda and RIKEN for support and hospitality while parts of the manuscript were finalised. The authors would also like to thank the Isaac Newton Institute
for hosting them at the programme "Applicable Resurgent Asymptotics", during the early stages of the work. IA
has been supported by the UK EPSRC Early Career Fellowship EP/S004076/1.

In addition the authors would like to thank Adri Olde Daalhuis, Fr\'ed\'eric Fauvet, Job Feldbrugge, Chris Howls, Omar Kidwai, Chris Lustri, Nikita Nikolaev, David Sauzin, Josh Shelton and Philippe Trinh for many interesting discussions. SC is particular grateful to Akira Shudo and Gerg\H{o} Nemes for many productive and enjoyable visits to Tokyo Metropolitan University where much of this work was undertaken.

\bibliographystyle{JHEP}
\bibliography{stokes}

\end{document}